\documentclass{aa}

\usepackage{afterpage}

\usepackage{graphicx}

\usepackage{newtxtext}
\usepackage[smallerops,cmbraces]{newtxmath}

\usepackage{xcolor}

\definecolor{xlinkcolor}{cmyk}{1,1,0,0}
\usepackage[
  bookmarks=true,         
  pdfnewwindow=true,      
  colorlinks=true,       
  linkcolor=xlinkcolor,  
  citecolor=xlinkcolor,  
  filecolor=xlinkcolor,  
  urlcolor=xlinkcolor,   
  final=true,
]{hyperref}

\usepackage{bm} 
\usepackage{textgreek}

\usepackage[capitalize]{cleveref}
\crefname{section}{Sect.}{Sects.}

\creflabelformat{equation}{#2#1#3}
\crefname{enumi}{item}{items}

\usepackage{siunitx}
\DeclareSIUnit[number-unit-product = ]\percent{\char`\%}

\usepackage{csquotes}

\definecolor{blackberry}{HTML}{8D1D75}

\DeclareSIUnit\parsec{pc}
\DeclareSIUnit\dex{dex}
\DeclareSIUnit\h{\mathnormal{h}}
\DeclareSIUnit\year{yr}
\DeclareSIUnit\years{yrs}
\DeclareSIUnit\arcsec{arcsec}
\DeclareSIUnit\arcmin{arcmin}
\DeclareSIUnit\Msun{M_\odot}
\DeclareSIUnit\Rsun{R_\odot}
\DeclareSIUnit\Lsun{L_\odot}
\DeclareSIUnit\Rvir{\mathnormal{R}_\mathrm{vir}}
\DeclareSIUnit\Rhalf{\mathnormal{R}_{1/2}}
\DeclareSIUnit\erg{erg}
\DeclareSIUnit\angstrom{\text{Å}}

\newcommand*{\Msun}{\ensuremath{\mathrm{M}_\odot}}
\newcommand*{\Rsun}{\ensuremath{\mathrm{R}_\odot}} 
\newcommand*{\Lsun}{\ensuremath{\mathrm{L}_\odot}} 
\newcommand*{\Rvir}{\ensuremath{R_\mathrm{vir}}}
\newcommand*{\Rhalf}{\ensuremath{R_{1/2}}}

\makeatletter

\renewcommand*\aa@pageof{, page \thepage{} of \pageref*{LastPage}}

\usepackage{natbib}
\bibpunct{(}{)}{;}{a}{}{,}
\def\bibfont{\aa@bibliographyfont}
\providecommand{\@LN}[2]{}

\makeatother

\newcommand{\rev}[1]{{\color{black} #1}}
\newcommand{\revb}[1]{{\color{black} #1}}

\begin{document}

\title{Cosmological zoom-in simulation of odd radio circles as merger-driven shocks in galaxy groups}
\titlerunning{ORCs in cosmological simulations}

\author{
    Anna Ivleva\inst{\ref{inst:usm}}
    \and
    Ludwig M. B\"oss\inst{\ref{inst:chi}}
    \and
    Klaus Dolag\inst{\ref{inst:usm},\ref{inst:mpa}}
    \and
    B\"arbel S. Koribalski\inst{\ref{inst:atnf},\ref{inst:wsu}}
    \and
    Ildar Khabibullin\inst{\ref{inst:usm},\ref{inst:mpa}}
}
\authorrunning{A. Ivleva et al.}

\institute{
    Universitäts-Sternwarte, Fakultät für Physik, Ludwig-Maximilians-Universität München, Scheinerstr. 1, 81679 München, Germany\label{inst:usm}\\
    \email{ivleva@usm.lmu.de}
    \and 
    Department of Astronomy and Astrophysics, The University of Chicago, William Eckhart Research Center, 5640 S. Ellis Ave. Chicago, IL 60637\label{inst:chi}
    \and
    Max-Planck-Institut für Astrophysik, Karl-Scharzschild-Str. 1, 85748 Garching, Germany\label{inst:mpa}
    \and
    Australia Telescope National Facility, CSIRO, Space and Astronomy, P.O. Box 76, Epping, NSW 1710, Australia\label{inst:atnf}
    \and 
    Western Sydney University, Locked Bag 1797, Penrith South DC, NSW 2751, Australia\label{inst:wsu}
}

\date{Received 01 August, 2025 / Accepted 09 December, 2025}

\abstract
{A new class of distinct radio objects, commonly referred to as odd radio circles (ORCs), has been recently discovered. The origin of these features remains unclear because their peculiar properties challenge our current understanding of astrophysical sources for diffuse radio emission.}
{We test the feasibility and limits of major mergers in galaxy groups as a possible formation channel for ORCs.}
{By modelling the assembly of a massive galaxy group with a final virial mass of $M_{200}\sim 10^{13}\,\Msun$ in a magnetohydrodynamic zoom-in simulation with on-the-fly cosmic ray treatment, we derive the X-ray and radio properties of the system self-consistently and compare them to observations.}
{We show that the X-ray properties of the simulated system agree with characteristics of observed galaxy groups in the relevant mass range, legitimating the comparison between the radio properties of the simulated halo and those of observed ORCs. A major merger between two galaxies in the simulation triggers a series of strong shocks in the circumgalactic medium, which in unison form a ring if the line of sight is perpendicular to the merger axis. The shock is rapidly expands radially and quickly reaches the virial radius of the halo. This formation channel thus readily explains the morphology and large extent of ORCs. However, the inferred radio luminosity of these features is lower than that of observed counterparts, while the degree of polarisation seems systematically over-predicted by the simulation.}
{Fossil cosmic ray populations from active galactic nuclei and stellar feedback might be necessary to explain the full extent of the radio properties of ORCs, since diffusive shock acceleration was the only source term for non-thermal electrons considered in this work.}


\maketitle

\section{Introduction}
\label{sec:introduction}

The current advent of next-generation radio telescopes delivers a number of interesting surprises to the astrophysics community. Their unprecedented depth and frequency coverage continue to reveal stunning details in various extraterrestrial settings—from galactic sources such as supernova remnants \citep[e.g.][]{Weiler:1986,Dokara:2021,Ball:2023,Do:2024} to the plentiful extragalactic zoo of objects such as star-forming disc galaxies \citep[e.g.][]{Martinsson:2016,Krause:2018,Stein:2023,Heesen:2024}, galaxies hosting active galactic nuclei \citep[AGN, e.g.][]{McKean:2021,Venturi:2022,Koribalski:2024c}, as well as haloes and radio relics of galaxy clusters \citep[e.g.][]{vanWeeren:2010,Rajpurohit:2020,Brueggen:2021,Venturi:2022,Cassano:2023,Koribalski:2024b}. However, it seems that yet another type of distinct source needs to be added to the list of possible radio objects. As their name suggests, odd radio circles (ORCs) are peculiar radio features with edge-brightened ring- or shell-like morphologies \citep{Norris:2021, Koribalski:2021, Koribalski:2024a}. The spectral indices measured in these features, that can lie between \mbox{$\alpha \approx 2$ and $0.6$} (assuming that the emission spectrum follows $I_\nu\rev{\,\propto\,}\nu^{-\alpha}$), suggest that these regions are filled with shock-accelerated cosmic rays (CRs) emitting via synchrotron at radio frequencies. \rev{In agreement with other current studies on this topic, we only consider objects at high galactic latitudes that exhibit an identified galaxy close to their geometric centre, in order to separate potentially different phenomena \citep{Koribalski:2024a,Norris:2025}.} Bona fide ORCs thus have a central early-type galaxy, strongly suggesting they are of extragalactic origin \citep{Koribalski:2021, Norris:2022, Norris:2025}. Since ORCs lack a counterpart in any other wavelength regime, distance and size measurements of ORCs rely on the redshifts of \rev{these} putative host galaxies. Judging by \rev{optical measurements of those objects, ORCs} lie at distances of $z=0.05-0.6$ and often have diameters of several hundred kiloparsecs, while the stellar masses of the central galaxies lie around $M_\ast=10^{11}\,\Msun$. This means that ORCs have enormous sizes compared to their central host, as their extent can even exceed the virial radius of the host system. \rev{While the majority of ORCs has been observed significantly above the Milky Way disc, we note that a few objects with similar morphologies have also been identified at low galactic latitudes \citep[e.g.][]{Koribalski:2024b}. But as mentioned above, they do not display a prominent central galaxy and differ in various other properties, indicating that these features are likely galactic supernova remnants and thus are formed in an entirely different setting \citep{Filipovic:2022,Omar:2022a,Sarbadhicary:2023}.}

Since the nature of ORCs remains under debate, careful control of the sample of considered objects is needed to ensure an appropriate comparison. The following set of conditions seems to \rev{work} best as a common denominator: the ORC must a) display a ring or shell-like morphology, b) host at least one elliptical galaxy in its centre, and  c) have a large radio ring size comparable to the virial radius of the central galaxy (group). While the neighbouring ORCs 2 and 3 from the original discovery paper by \citet{Norris:2021} — who first reported such radio morphologies and have introduced the term ORC — \rev{likely have an AGN-related nature \citep[e.g.][]{Norris:2021b,Norris:2025,Shabala:2024}}, the origin of a growing list of different ORCs remains unclear. It is \rev{also} important to note that the observed morphologies can depend on the resolution and frequency of the observation facility. While an object may look diffuse in one image, the characteristic ring emission for ORCs can reveal itself in surveys with higher resolution and sensitivity.

The following observed objects are falling under the \rev{declared criteria}: ORCs~1 and 4 \citep{Norris:2021}, ORC~5 \citep{Koribalski:2021}, ORC~6 \citep{Dolag:2023, Koribalski:2025}, \mbox{ORC~J2223--4834} \citep{Gupta:2022}, SAURON \citep{Lochner:2023}, Physalis \citep{Koribalski:2024a}, Cloverleaf \citep{Bulbul:2024}, \mbox{ORC~J1027--4422} \citep{Koribalski:2024b}, \mbox{ORC~J0219--0505} \citep{Norris:2025}, \rev{\mbox{ORC J0356–4216} \citep{Taziaux:2025} and \mbox{RAD J131346.9+500320} \citep{Hota:2025} The benchmarks stated above for ORCs are rather restrictive and thus some sources published as ORCs are not included in our list. Nevertheless it is difficult to draw a definite line for candidates, as one cannot be sure of the exact set of properties they need to display\footnote{\rev{e.g. the diameters of Physalis \revb{(\mbox{145 kpc})} and ORC J0219–0505 \revb{\mbox{(114 kpc)}} are relatively small compared to the other candidates \revb{(a few hundred kiloparsecs)}. Still the extent of their radio features is \revb{large} compared to galaxy scales, and since all other properties fit very well with the rest of observed ORCs, we include them in our list.}}. We therefore build a quite conservative list, attempting to group only those objects likely to have a similar formation path. Nevertheless we note the several ongoing searches for ORC-like systems, as the study by \citet{Gupta:2025}.}

A variety of extragalactic formation scenarios for ORCs have been proposed. Several studies have addressed the scenario of a termination shock due to galactic outflows during starburst events \citep{Norris:2021,Norris:2022,Coil:2024}. Another possibility is that a black hole drives these features via binary mergers of supermassive black holes \citep{Koribalski:2021,Norris:2022}, tidal disruption of stars by black holes \citep{Omar:2022b}, or AGN jets, where the ejected radio lobes can appear circular under certain projections and physical conditions \citep{Nolting:2023,Fujita:2024,Lin_Yang:2024}. In this regard, \citet{Shabala:2024} have pointed out that the radio power of a mere AGN jet would decline too quickly to explain the properties of ORCs and hence concluded that an additional shock running through pre-existing AGN bubbles would be necessary in this scenario. Finally, there is the shock-driven formation channel first proposed by \cite{Dolag:2023} and \cite{Koribalski:2024a}. Based on cosmological simulations, the suggested concept posits that merger-accelerated shocks -- i.e. merger shocks from coalescence events in massive galaxies and groups encountering outer accretion shocks -- can produce features that closely resemble ORCs. Such processes have been studied for a while now in relation to galaxy clusters, where merger shocks are most likely \rev{the reason} for radio relics \citep[e.g.][]{Vazza:2009,Ha:2018,Zhang:2019}. But in the case of clusters, these features always lie well within the boundaries of the respective halo. Hence, ORCs could be unprecedented evidence for merger shocks in galaxy and group haloes, where their extent pushes beyond the virial radius. However, the simulations by \cite{Dolag:2023} predicted radio luminosities that were too low compared to observations, concluding that the relatively low simulated halo mass of $10^{12}\,\Msun$ could be a major issue, since the total energy budget is underestimated. This problem was also pointed out by \citet{Yamasaki:2024}, who analytically explored the feasibility of pure accretion shocks as the source for ORCs and concluded that more complex simulations with a larger halo mass of at least $10^{13}\,\Msun$ are required matching the estimated halo masses of ORC host galaxies \rev{\citep{Norris:2021b,Koribalski:2024a}}.

In the present work, we address the need for such a study. We present a first-of-its-kind zoom-in simulation of a galaxy merger in high resolution, where the final group mass corresponds to observed haloes in ORCs. We performed a non-radiative magnetohydrodynamic simulation with an on-the-fly treatment of spectral CR electrons. First, we summarise the numerical details of the simulation by stating the important features of the code and initial condition in \cref{sec:code}. In \cref{sec:results}, we give a detailed presentation of the simulation results. We outline the time evolution of the merger and then describe the halo's X-ray and radio properties, including inferred mock observation images. Particular focus is placed on the polarisation of the simulated ORC. Similarities and differences with observations are discussed in \cref{sec:discussion}, before we summarise our results in \cref{sec:summary}.

\section{Simulation code and setup}
\label{sec:code}

\begin{table}[!t]
\caption{Key properties of the simulation. $m_{\mathrm{dm}}$ and $m_{\mathrm{gas}}$ denote the mass resolution of the dark matter and gas particles, respectively, while $\epsilon_{\text {dm}}$ and $\epsilon_{\text {gas}}$ indicate the gravitational softening length used for corresponding particle species in the simulation. The value $h=H_{0}/\mathrm{(100~km~s^{-1}~Mpc^{-1})}$ denotes \rev{the} reduced Hubble parameter.}
$$
{\begin{array}{cccc}
\hline \hline \begin{array}{c}
m_{\mathrm{dm}}\\
\left[h^{-1} \rm M_{\odot}\right]
\end{array}& \begin{array}{c}
m_{\mathrm{gas}} \\
\left[h^{-1} \rm M_{\odot}\right]
\end{array} & \begin{array}{c}
\epsilon_{\text {dm }} \\
\left[h^{-1} ~\mathrm{kpc}\right]
\end{array} & \begin{array}{c}
\epsilon_{\text {gas }} \\
\left[h^{-1} ~\mathrm{kpc}\right]
\end{array}\\[3mm]
4.0 \times 10^5 & 0.6 \times 10^5 & 0.36 & 0.12 \\
\hline
\end{array}}
$$
\label{tab:simprops}
\end{table}

\begin{figure*}[!t]
\centerline{
    \includegraphics[width=\textwidth, trim={0 0 0 0}]{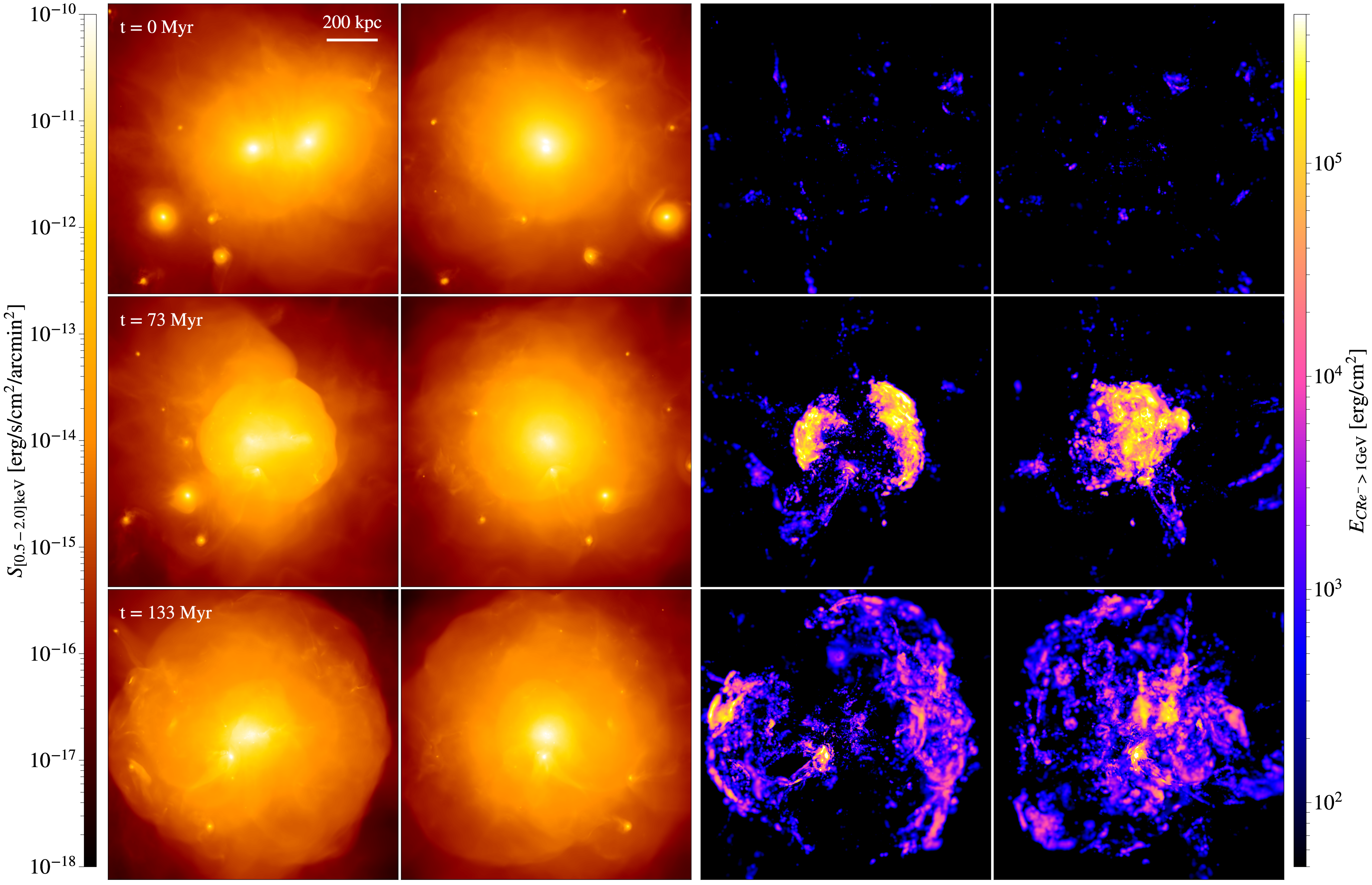}}
    \caption{Time evolution of the circumgalactic medium of the simulated galaxy group, showing the central region at three successive moments close to the merger event. The two columns display maps of the X-ray surface brightness and the projected energy density of cosmic ray electrons, while the two panels in each column represent two orthogonal sight lines. For the X-ray map, we assume a distance of $z=0.017$. The scale bar (the top left panel) indicates \SI{200}{\kilo\parsec}. The time labels provide the elapsed time since shortly before the merger event.}
    \label{fig:Xray_CRe}
\end{figure*}

The simulation was carried out with the cosmological code \mbox{\textsc{OpenGadget3}}, which utilises a smoothed particle magnetohydrodynamics (SPMHD) scheme to model gas dynamics and a Tree-PM solver for gravity (release paper \rev{by Dolag et al.} in prep.). Its infrastructure is based on its predecessor Gadget2 \citep{Springel:2005} but employs major improvements in the gravity solver \citep{Ragagnin:2016} and SPH implementation \citep{Beck:2016a}, where also a revised time step criterion ensures the stability of the code around strong shocks in a simulation \citep{Pakmor:2012}. We used the Wendland C4 function as the kernel to reconstruct the local hydrodynamic properties of the simulated fluid based on the position and smoothing length of the simulated particles \citep{Dehnen:2012,Donnert:2013}. The code includes both a physically motivated conductivity \citep{Jubelgas:2004,Dolag:2004,Petkova:2009}, as well as an artificial conduction and time-dependent viscosity scheme \citep{Price:2012,Beck:2016a}, which provides numerical convergence when compared to other hydrodynamic discretisation methods.

The code resolves the magnetohydrodynamic equations in Lagrangian form \citep{Dolag:1999,Dolag:2009} in the non-ideal regime, i.e. with non-vanishing resistivity \citep{Bonafede:2011}. Non-zero $\nabla \cdot \mathbf{B}$ contributions from the magnetic field $\mathbf{B}$, arising from numerical truncation errors, are suppressed by the hyperbolic divergence cleaning method described by \citet{Tricco:2016}, whose implementation in \textsc{OpenGadget3} is described by Steinwandel and Price (in prep.). We ran the code with an on-the-fly shock finder \citep{Beck:2016b}, that returns the shock's hydrodynamic quantities, as well as the Alfv\'enic Mach number and shock obliquity \citep{Boess:2023b}. Finally, we includes CR electrons with the module CRESCENDO, a Fokker-Planck solver for spectral CRs \citep{Boess:2023b}. The sub-grid model computes and evolves the distribution function of CRs for each resolution element (i.e. SPH particle) at each time step in the simulation. The source for CRs is diffusive shock acceleration (DSA), which heavily relies on the output of the built-in shock finder and uses shock-obliquity-dependent acceleration efficiencies \citep{Pais:2018,Kang:2024}. These injected populations evolve further over time through adiabatic losses as well as radiative losses, the latter being inverse Compton scattering and synchrotron emission for CR electrons \citep[for the details on the computation of the synchrotron emissivity, see Appendix 2 by][]{Boess:2023}. The present simulation was performed in a non-radiative regime without star formation, resolving only the dark matter and gas evolution.

\begin{figure*}[!t]
\centerline{
    \includegraphics[width=\textwidth, trim={0 0 0 0}]{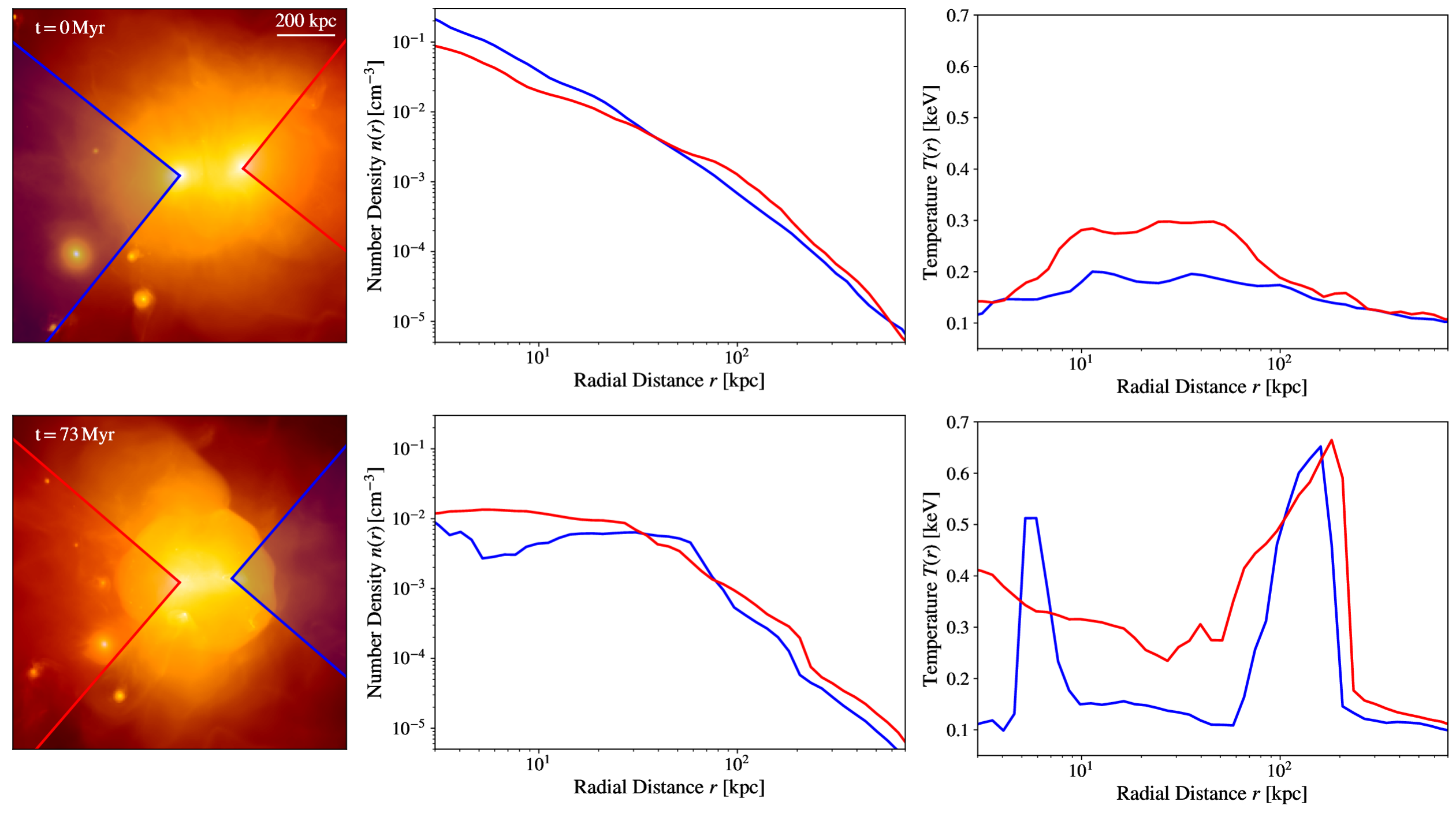}}
    \caption{Radial profiles of the hot X-ray emitting gas properties in the circumgalactic medium of the merging galaxies. Left, from top to bottom: X-ray surface brightness maps shortly before and after the merger event (see \cref{subsec:xray_CReEnergy}). The blue- and red-shaded regions in each panel indicate the extent of the 3D cones used to compute the radial profiles of the gas number density $n(r)$ (centre column) and temperature $T(r)$ (right column) for the two merging galaxies. The cones are centred on the local potential minimum of the respective dark matter halo.}
    \label{fig:CGM_props}
\end{figure*}

The initial condition for the simulation was drawn from the COMPASS\footnote{\url{http://www.magneticum.org/complements.html\#Compass}} set. It is a suite of zoom-in setups for a diverse selection of haloes -- from massive clusters down to Milky Way-like galaxies -- selected from a dark matter-only cosmological simulation with a box length of \SI{1}{\giga\parsec} \citep[e.g.][]{Bonafede:2011}. We chose the \textit{dfrogin} region, which produces a galaxy group halo with a final virial mass of $M_{200,\rm c}=1.8 \cdot 10^{13}\,\Msun$ \citep{Schlachtberger:2014}. \rev{The mass resolution for the gas and dark matter particles, as well as the corresponding smoothing lengths in the simulation are listed in \cref{tab:simprops}.} The simulation employs the standard \mbox{WMAP-7 $\Lambda$CDM} cosmological parameters \citep{Komatsu:2011}.

\section{Results}
\label{sec:results}

Until $z=0$, the halo undergoes an assembly process that includes various mergers with different galaxies. Shortly after $z=1$, the main halo experiences a particularly violent event, with a mass ratio of $1:2.4$. It is this major merger that triggers the shock and CR injection with subsequent radio emission that we identify as an ORC in this paper. In the following sections, we analyse the time evolution and properties of the galaxy group around this merger event. \rev{Where required, we adopt the distance to ORC Physalis while creating the mock maps and compare simulated properties to that observed counterpart. We acknowledge that it is not necessarily a representative case for the overall population of ORCs. However, given that it is found in the smallest observational volume with the least stringent observational constraints imposed by survey sensitivity, angular resolution and searching techniques, we consider it as a prototype for a large fraction of the ORC population\footnote{\rev{Physalis is the closest observed ORC so far and it would be highly unlikely to find a special case of an already rare phenomenon in such a small observational volume. This motivates the hypothesis that it represents a typical case, since ORCs with comparably small sizes and radio luminosities would not be detected at the much larger distances where other ORCs have been found.}}. Therefore, Physalis offers the best opportunities for a comparison between observational data and simulations at the moment.}

\subsection{Gas and shock morphology}
\label{subsec:xray_CReEnergy}

\cref{fig:Xray_CRe} shows the evolution of the system around the time of the major merger event. The rows represent three different output times separated by $\sim$\SI{70}{\mega\year}. The left and right halves of the figure display the X-ray surface brightness in the $[0.5-2.0]\,\rm keV$ band and the CR electron energy density, respectively. The X-ray flux was estimated based on contributions from thermal and line emission (assuming $Z=0.3\,\rm Z_\odot$), using the software \textsc{Smac} \citep{Dolag:2005}. We assumed a distance of \SI{75}{\mega\parsec} (corresponding to $z=0.017$), as for the observed ORC Physalis \citep{Koribalski:2024a}. We show two projections in each case, \rev{displaying} the system orthogonal and parallel to the merger axis in the left and right panels, respectively. For the rest of this work, we adopt $t=0$ as the time shortly before the merger, visualised in the first row of \cref{fig:Xray_CRe}. Almost directly after impact, two large shock shells emerge from the centre, which together form an almost full circle with a diameter of $300 - 400 \, \rm kpc$ after \SI{\sim70}{\mega\year} (third panel in second row of \cref{fig:Xray_CRe}). While surpassing the diameter of ORC Physalis, these dimensions generally agree well with the sizes of other observed ORCs, since Physalis is actually on the lower-size end \citep[cf. Table 3 by][]{Koribalski:2024a}. The clear boundary, where the X-ray brightness drops, coincides with the position of the shock front, traced by the CR electron population in the simulation. The total magnitude of the X-ray flux generally agrees with observations of haloes in this mass range \citep{Bulbul:2024}. Meanwhile, the merger shock further expands in radial direction, reaching a diameter of \SI{\sim1}{\mega\parsec} about \SI{130}{\mega\year} after the merger event, hence far surpassing the virial radius of the system equal to \SI{\sim440}{\kilo\parsec}.

\subsection{X-ray properties of the circumgalactic medium}
\label{sec:xray_props}

\begin{figure*}[!t]
\centerline{
    \includegraphics[width=\textwidth, trim={0 0 0 0}]{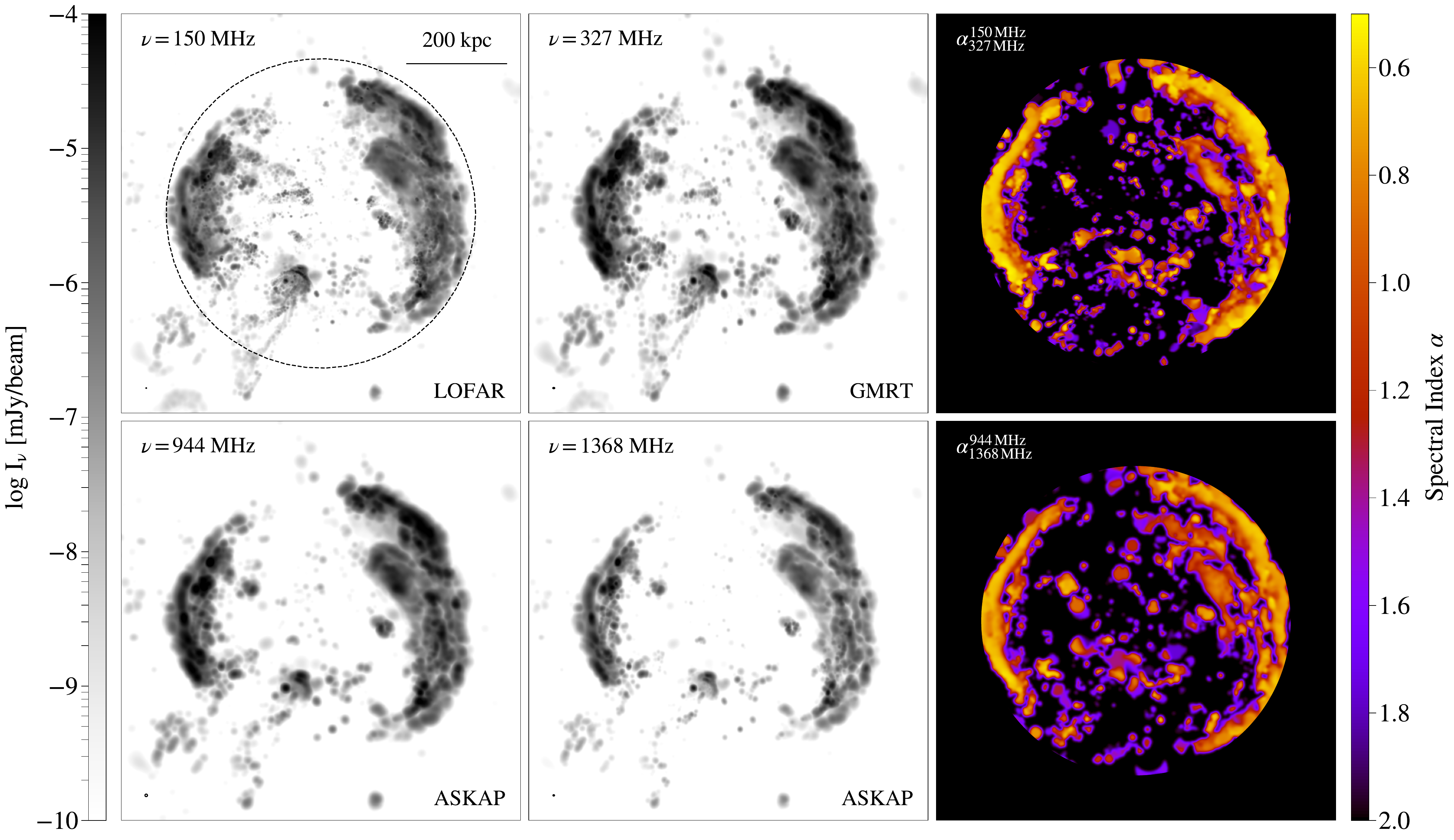}}
    \caption{Mock images of the simulated ORC at $t=\SI{73}{\mega\year}$. Left and centre: Synchrotron flux density at four frequencies (indicated in the top-left corner of each panel). The images are smoothed with a Gaussian kernel length matching the respective resolution of the telescope, shown in the bottom of each panel. Right: Spectral index derived from the two left panels. We applied a circular mask in these images to highlight the circular shape, where its extent is indicated by the dashed line in the upper left panel.}
    \label{fig:synch_spectralindex}
\end{figure*}

First, we focus on the evolution of the hot gas component and its inferred X-ray properties. \cref{fig:CGM_props} demonstrates them for two different times -- shortly before and after the merger. The first column displays the same X-ray maps as in the first two rows of \cref{fig:Xray_CRe} in the orthogonal projection, while the second and third column show the radial profiles of the mass-weighted median of the electron density $n(r)$ and temperature $T(r)$ for the gas of both merging haloes, respectively. Note that as a filter for the X-ray emitting gas, we applied density and temperature cuts of \mbox{$n<0.3 \,\rm cm^{-3}$} and \mbox{$T>10^5\, \rm K$}, respectively. After the first pericentre passage, the gaseous component of the galaxies was effectively shredded, while the dark matter component maintains two significant overdensities. Therefore, we centred the profiles on the two potential minima inferred from the underlying dark matter distribution, which mixes on longer timescales. To minimise contamination by the other halo, the radial profiles were only computed in 3D cones with an opening angle of \SI{90}{\degree}, pointing away from each other. These cones are indicated by the red- and blue-shaded regions in the X-ray map on the left in \cref{fig:CGM_props}, where the respective colour is also adopted for both galaxies in the adjacent radial profiles.

Before the merger, at \mbox{$t=0$}, the density and temperature profiles of both galaxies suggest two relaxed galaxies, where the gas is in hydrostatic equilibrium. The central density, and hence total mass of one halo (blue line), is approximately two times higher than the other (the exact merger mass ratio is \mbox{$1:2.4$}). At \mbox{$t=\SI{73}{\mega\year}$} -- after the initial pericentre passage -- the gas properties exhibit considerable differences in both density and temperature. While the initially less massive galaxy (red line) develops a pronounced core, the other displays a significant drop in density at its inner few kiloparsecs, corresponding to a peak in temperature for this galaxy. Both haloes exhibit a large temperature jump around a radial distance of \SI{100}{\kilo\parsec}, after which the temperature in the circumgalactic medium (CGM) of both galaxies drops again. This coincides with the position of the shock at this moment, also evident by the drop of the X-ray flux visible in the map on the left of \cref{fig:CGM_props}. The pericentre passage led to massive stripping of gas from the central regions of both galaxies, visible by the strong decrease in density compared to the picture before. However, one galaxy (red line) maintains a much denser core compared to the other one, whose central density is almost an order of magnitude lower.

This has interesting consequences for the X-ray brightness of both galaxies. The gas luminosity due to thermal X-ray emission is only weakly dependent on temperature while being quite sensitive to gas density. Since one galaxy here (red line) has a significantly higher central density, it consequently outshines the second galaxy. X-ray observations of ORC Physalis revealed that the X-ray flux of the system is centred on one of its two merging galaxies, while the other does not display any significant central emission \citep[cf. Fig. 2 by][]{Koribalski:2024a}. Our simulation managed to reproduce this behaviour, considering that the galaxy densities overall remain very low after the merger. This could explain why one halo is still detected, while the other is too dilute to be visible in X-ray. Forthcoming deeper X-ray data accumulated for Physalis by XMM-Newton (Khabibullin et al., in prep) should allow one to test this prediction.

\subsection{Mock radio observations}
\label{subsec:radiomocks}

In \cref{fig:synch_spectralindex} we present mock images of the radio flux density due to synchrotron emission at four different frequencies and the resolved spectral indices of the simulated ORC. The orientation of the sight line is orthogonal to the merger axis, and we chose the time of best morphological resemblance to observed counterparts ($\sim$\SI{70}{\mega\year} after the merger event). We shifted the object from the simulation redshift of $z=0.3$ to $z=0.017$, in order to be consistent with the distance to the observed ORC Physalis \citep{Koribalski:2024a}\footnote{Strictly speaking, this slightly affects the inferred radio properties of the system, since we include redshift-dependent energy losses of CRs via inverse Compton scattering with photons of the cosmic microwave background. The spectra are therefore biased towards higher spectral indices here. Nevertheless, the \rev{qualitative} changes to the results are insignificant.}. The frequencies and Gaussian smoothing kernel widths were chosen such that they coincide with those of currently operating radio telescopes: \SI{150}{\mega\hertz} ($\theta=\SI{3.3}{\arcsecond}\times\SI{3.3}{\arcsecond}$, LOFAR), \SI{327}{\mega\hertz} ($\theta=\SI{11.4}{\arcsecond}\times \SI{9.8}{\arcsecond}$, GMRT), \SI{944}{\mega\hertz} ($\theta=\SI{15}{\arcsecond}\times\SI{15}{\arcsecond}$, ASKAP) and \SI{1368}{\mega\hertz} ($\theta=\SI{8.7}{\arcsecond}\times\SI{8.0}{\arcsecond}$, ASKAP). The respective telescope and observer frequency is written in each panel, while the scale in the upper left panel indicates a physical size of \SI{200}{\kilo\parsec}. Similarly to other current studies, our simulation underestimates the local magnetic field strength since we lack the ultra-high resolution necessary to resolve the turbulent dynamo and hence the main driver for efficient magnetic field amplification \citep{Steinwandel:2020}, as well as astrophysical seeding mechanisms \citep[e.g.][]{Beck:2013,Garaldi:2021}. In order to provide realistic mock images of the synchrotron brightness -- which in turn heavily relies on the magnetic field strength -- we calculated the synchrotron emissivity by assuming equipartition between turbulent and magnetic pressure \rev{throughout this work}. This assumption implies a magnetic field strength of $B^2 = 4\pi\rho v^2_{\rm turb}$, where $\rho$ and $v^2_{\rm turb}$ are the local gas density and its turbulent velocity component, respectively \citep[][see \cref{sec:discussion} for details]{Zhou:2024,Boess:2024}. Recent simulations indicate that this saturation of the turbulent dynamo is valid in regions with strong shocks, where the generation and amplification of magnetic fields is particularly efficient \citep{Steinwandel:2024,Zhou:2024}.

The diffuse structure visible at low frequencies is successively reduced at increasing frequencies. This is due to the strong energy dependence of the CR cooling time. Therefore, we see high-energy CR electrons predominantly only at shocks, where they are currently accelerated. Electrons at lower energies, however, maintain their energy longer after being shock-accelerated and hence continue to emit in radio downstream of the shock. This explains the difference between the two spectral index maps $\alpha_{327\rm\, MHz}^{150 \rm\, MHz}$ and $\alpha_{1368\rm\, MHz}^{944\rm\, MHz}$ in the right column of \cref{fig:synch_spectralindex}, which indicate the spectral slope of the radio flux density at low and high frequencies, respectively ($I_\nu \propto\nu^{-\alpha}$). In these images we applied a radial mask, in order to reduce the noise in the outskirts (indicated by the dashed circle in the upper-left panel). The spectral slope at low frequencies is generally flatter (with maximum values of about $0.6$ at the shock front), than at high frequencies (about $0.9$ at shock), where the bulk of CR electrons has already cooled off to lower energies, causing a steepening of the spectrum. This aspect, as well as a comparison to observed counterparts, is described in \cref{subsec:spectrum_radiopower}.

\subsection{Emission spectrum and total radio power}
\label{subsec:spectrum_radiopower}

\begin{figure*}[!t]
\centerline{
    \includegraphics[width=0.5\textwidth, trim={0 0 0 0}]{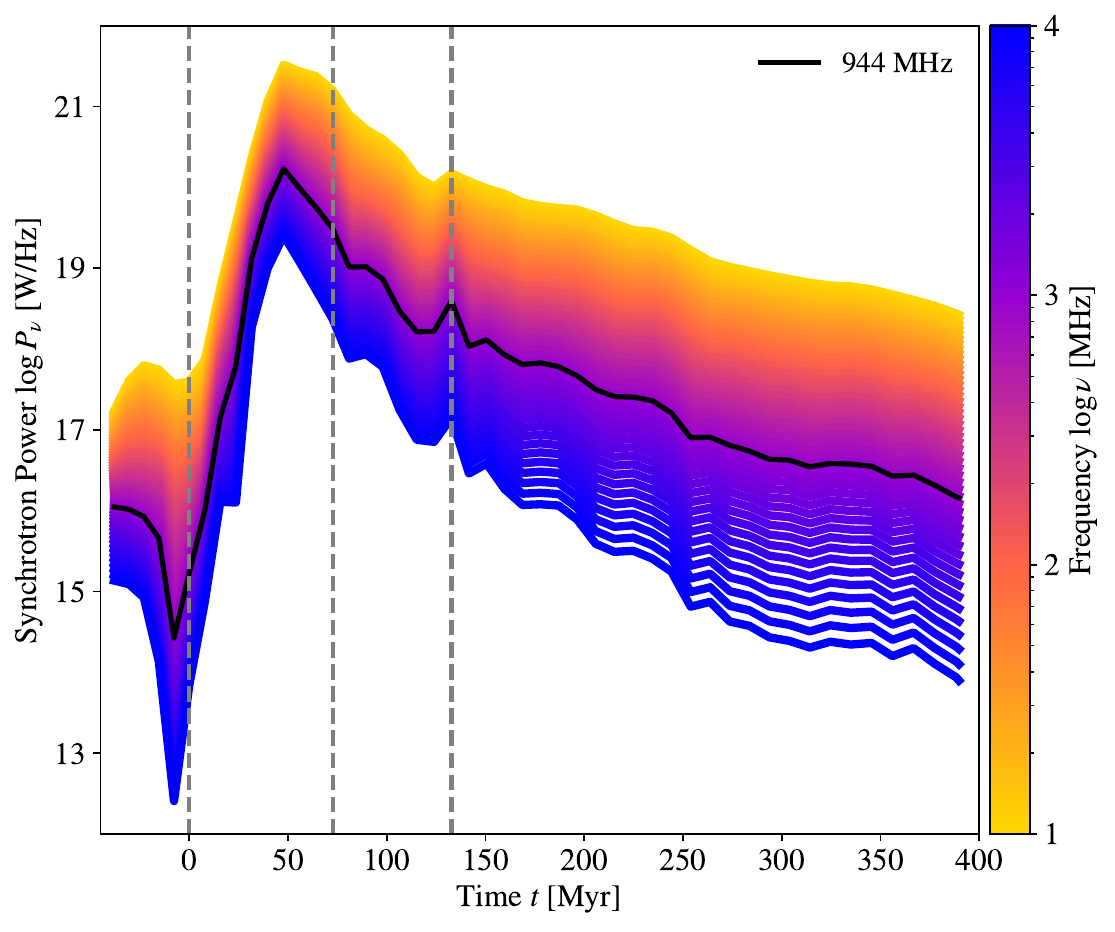}
    \includegraphics[width=0.515\textwidth, trim={0 0 0 0}]{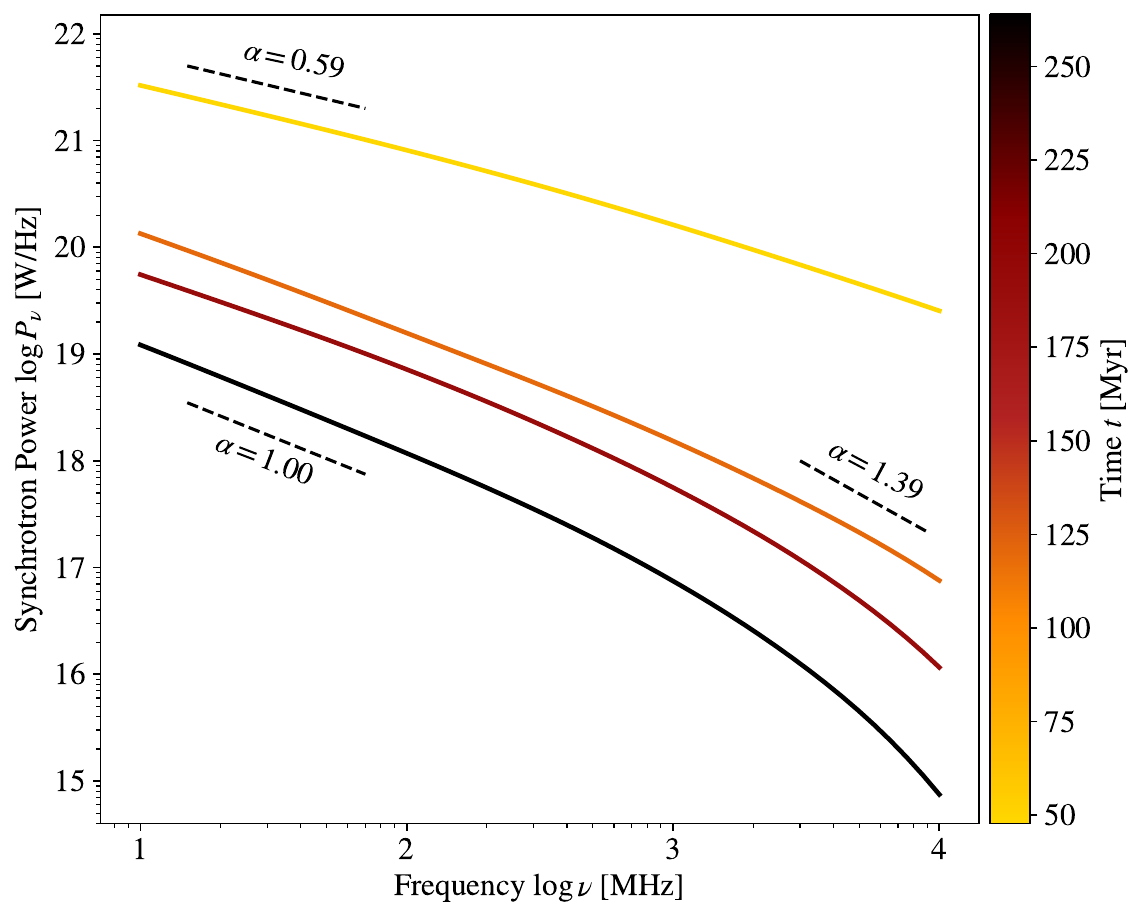}}
    \caption{Total synchrotron power and spectrum $P_\nu$ over time $t$. Left: Time-dependent luminosity profile for varying frequencies $\nu$. The solid black line represents $\nu=\SI{944}{\mega\hertz}$. The vertical dashed lines indicate the three time steps at which the simulated ORC was visualised in \cref{fig:Xray_CRe}. Right: Associated emission spectrum for four equidistant time steps. Here, the dashed lines indicate the local spectral index according to $P_\nu\propto\nu^{-\alpha}$.}
    \label{fig:radiopower_spectrum}
\end{figure*}

According to the formation model analysed in this paper, the radio feature that we identify as an ORC is a merger shock around a massive galaxy or galaxy group, triggered along the axis of a major merger. In this scenario, the total radio luminosity would be the highest shortly after the merger event, since that would be the time when the shock can inject relatively many CR electron populations efficiently into the dense medium at small radii. If we assume that CRs get accelerated from the thermal pool according to DSA, the \rev{power-law} index $\gamma$ of the injected CR energy spectrum depends only on the Mach number $\mathcal{M}$. Rewriting $\gamma$ in terms of the spectral index $\alpha$ of the emission spectrum yields

\begin{equation}
    \alpha = \frac{\mathcal{M}^2+3}{2\mathcal{M}^2-2}. \label{eq:spectralindex_mach}
\end{equation}

\noindent In the limit of a strong shock ($\mathcal{M}\rightarrow\infty$), the spectral index is $\alpha\rightarrow1/2$, while weaker shocks generate steeper spectra ($\alpha>1/2$). After the initial central merger, the shock travels radially outwards with decreasing effective Mach numbers, producing increasingly steep spectra at the front.

In \cref{fig:radiopower_spectrum} we present the time evolution of the ORC's total luminosity produced by synchrotron emission. For this purpose, we extracted the particles around the expanding shock front at each time step and calculated their cumulative radio brightness at 50 frequencies equidistant in $\log_{10}$ space according to Eq. A1 by \citet{Boess:2023}. The two panels show the total brightness on the y-axis, while the x-axis represents time $t$ and frequency $\nu$ in the left and right panel, respectively. On the left, different colours identify varying frequencies $\nu$ according to the colour bar, while the solid black line highlights specifically $\nu=\SI{944}{\mega\hertz}$, namely the frequency at which ORC Physalis \citep{Koribalski:2024a} was observed. The vertical dashed lines indicate the three times at which the simulation was displayed in \cref{fig:Xray_CRe} ($t=0, 73, \SI{133}{\mega\year}$). Shortly after $t=0$, the merger triggers the shock -- evident by the sudden jump in luminosity -- and reaches its maximum by around $t=\SI{50}{\mega\year}$. Afterwards, the total power steadily decreases in all frequencies, with the decline being stronger at higher frequencies. This effect becomes even more apparent on the right side of \cref{fig:radiopower_spectrum}. This panel shows the cumulative emission spectrum of the simulated ORC at four successive times, which are approximately $\SI{70}{\mega\year}$ apart, starting from the time of maximum radio brightness at $t=\SI{47}{\mega\year}$ (yellow line). Early on, we see a relatively flat spectral index of $\alpha\approx0.59$ in the lower frequency range. Moving on to later times, the spectrum rapidly decreases in total luminosity, with further spectral steepening at higher frequencies. This is due to two effects: while the shock travels outwards, its effective Mach number decreases, resulting in a lower acceleration efficiency in the DSA framework implemented in our code. Newly injected CRs therefore display increasingly steep spectra with time (see \cref{eq:spectralindex_mach}), while simultaneously the \rev{normalisation} of the spectrum systematically decreases. The latter feature is also influenced by the decline in local gas density and magnetic field strength with radius, since these circumstances lead to less gas with new CR populations and lower synchrotron emissivity, respectively. For a fixed time step, the steeper slope at high frequencies is caused by radiation losses, since the energy loss timescale of electrons due to Synchrotron emission and inverse Compton scattering strongly depends on their energy. This leads to more rapid energy loss of CR electrons at higher energy bins, which is reflected in the steeper emission spectrum at higher frequencies, since the contribution to the spectral emission of a charged particle in a magnetic field is closely peaked around an energy-dependent maximum \citep{Ginzburg:1979}.

Taking the spectral index at the earliest time ($\alpha=0.59$) as a proxy for the overall maximum over time, we can easily estimate the respective Mach number $\mathcal{M}$ by rearranging \cref{eq:spectralindex_mach} for $\mathcal{M}$. \rev{This would yield a radio-inferred Mach number of $\mathcal{M}\approx4.8$ at the ORC, which is comparable to measurements of merger events in galaxy clusters. The actual maximum of the Mach number distribution is lower (around $\mathcal{M}\approx3$). However, the distribution is relatively broad (see \cref{fig:machdistribution} in the Appendix).} The maximum radio power at $\nu=\SI{150}{\mega\hertz}$ is $P_\nu=9.3\times10^{19}\SI{}{\,\watt/\hertz}$, which is about three orders of magnitude lower than the measurement for ORC Physalis \citep{Koribalski:2024a}.

\subsection{Polarisation}
\label{subsec:polarization}

Synchrotron radiation from electrons is polarised with a polarisation angle $\chi$, i.e. the angle between the electric field vector of the associated radiation, and a chosen unit vector in the sky plane, defined as orthogonal to the line of sight. Since the polarisation angle is \rev{only dependent} on local properties, the so-called polarisation fraction $\pi_\nu$ is relatively high in case of a singular, relatively small emitting CR cloud where the local magnetic field is approximately uniform. But as the number of contributing electrons along the line of sight increases, the fractional polarisation of the observed radiation consequently decreases because the direction of the magnetic field is not homogeneous. Therefore, a polarisation analysis of radio observations can be a powerful tool to dissect the physical processes relevant in ORCs. Since our simulation self-consistently evolves CR electrons \rev{and the magnetic field direction}, we can provide polarisation information of the simulated ORC and compare it with observed counterparts for the first time.

The polarisation state of electromagnetic radiation at a given frequency $\nu$ is commonly described by the four Stokes parameters $I_\nu, Q_\nu, U_\nu$ and $V_\nu$, since they can be inferred directly from observed quantities. Generally, synchrotron radiation is elliptically polarised. But in the limit of ultra-relativistic electrons, the light cone of radiation -- defined as the solid angle around the velocity vector within which an observer receives a significant signal -- becomes \rev{much smaller than $4\pi$}. In this case, the observer receives only 'flashes' of synchrotron radiation when the electron moves almost exactly towards the observer. The associated electric field vector hence only draws a line in the sky plane (instead of an ellipse), which is equivalent to linear polarisation \citep{Ginzburg:1979,Longair:2011}. In this limit, $V_\nu$ is vanishes and only three Stokes parameters remain. Assuming the line of sight is along the $z$-direction, $I_\nu$ is defined as the total intensity received by the observer integrated along the line of sight:

\begin{equation}
    I_\nu = \int j_\nu(\mathbf{x}) \, dz,
\end{equation}

\noindent where $j_\nu(\mathbf{x})$ is the synchrotron emissivity at a given frequency $\nu$ and position $\mathbf{x}=(x,y,z)$. Meanwhile $Q_\nu$ and $U_\nu$, which carry the polarisation information, are defined by

\begin{align}
    Q_\nu &= \int j_{\nu,\rm pol}(\mathbf{x}) \cos2\chi \, dz\\
    U_\nu &= \int j_{\nu,\rm pol}(\mathbf{x}) \sin2\chi \, dz,
\end{align}

\noindent where $j_{\nu,\rm pol}$ is the polarised synchrotron emissivity. Note that the polarisation angle \rev{$\chi_0$} is subject to Faraday rotation along the line of sight. The polarisation angle therefore needs to be adjusted at each contributing particle along the line of sight according to the travelled distance $dl$ by

\begin{equation}
    \rev{\chi = \chi_0} + \lambda^2 \text{RM} \quad \text{with} \quad \text{RM}= \rev{0.812 \, \text{rad} \, \text{m}^{-2}} \int n_e B_z \, dl, \label{eq:faradayrotation}
\end{equation}

\noindent where $\lambda$ is the observed wavelength, RM the rotation measure, $n_e$ the local electron number density and $B_z$ the component of the magnetic field vector $\mathbf{B}=(B_x, B_y, B_z)$ along the line of sight. The polarisation angle depends only on the magnetic field components in the sky plane and is given by

\begin{equation}
    \cos2\rev{\chi_0} = \frac{B_x^2 - B_y^2}{B_x^2 + B_y^2}.
\end{equation}

\noindent Finally, the polarisation fraction $\pi_\nu$ can be obtained via

\begin{figure}[!t]
\centerline{
    \includegraphics[width=0.5\textwidth, trim={0 0 0 0}]{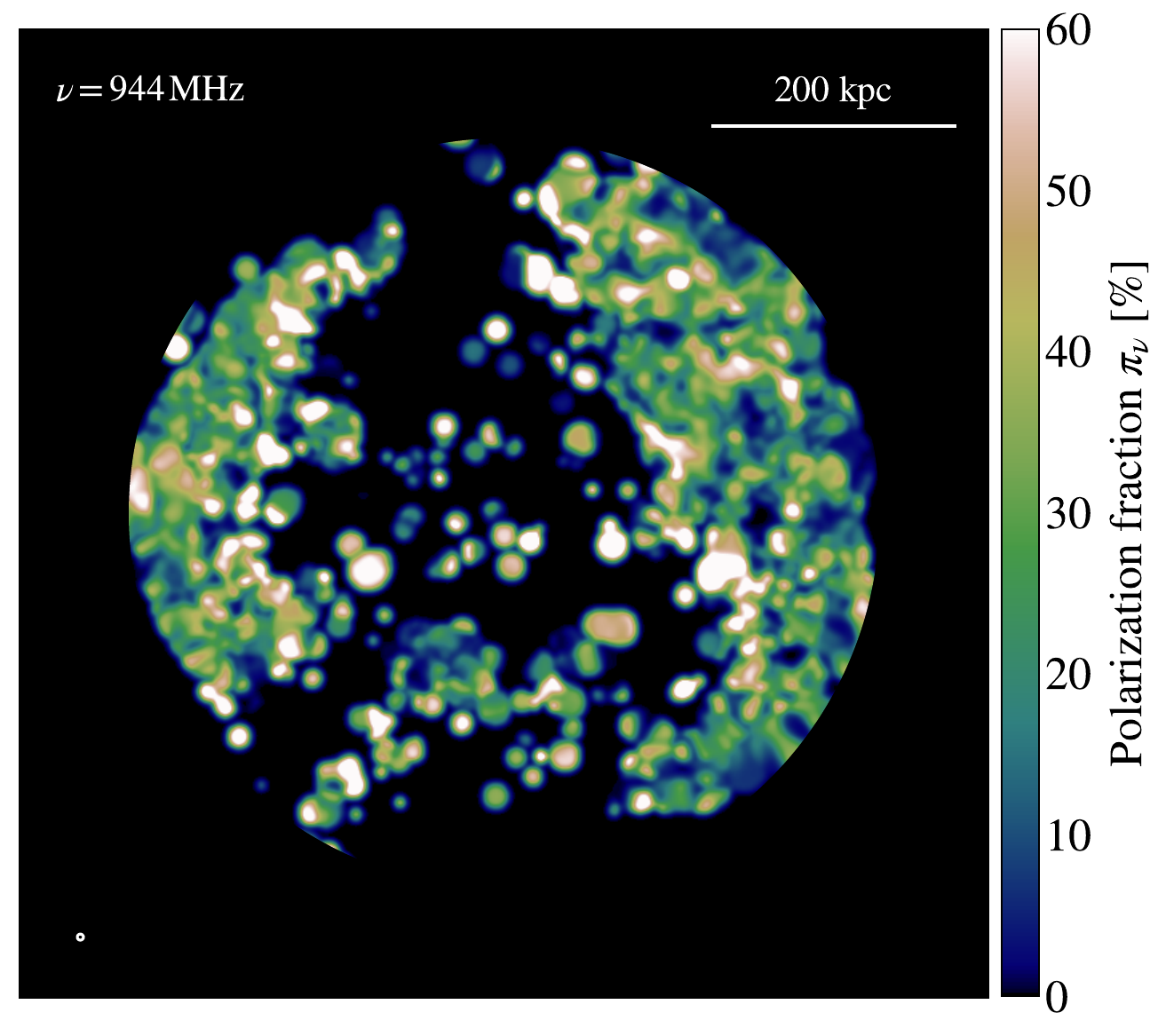}}
    \caption{Polarisation fraction $\pi_\nu$, as given by \cref{eq:polfrac} for $\nu=\SI{944}{\mega\hertz}$ at $t=\SI{73}{\mega\year}$. The Stokes parameters were smoothed according to the resolution of ASKAP at the frequency indicated in the bottom left corner (see text for details). The horizontal line in the top right corner displays a size of \SI{200}{\kilo\parsec}.}
    \label{fig:polarization}
\end{figure}

\begin{equation}
    \pi_\nu = \frac{\sqrt{Q^2_\nu + U^2_\nu}}{I_\nu}.\label{eq:polfrac}
\end{equation}

\noindent Generally, two factors will influence the degree of polarisation of synchrotron radiation: i) the energy spectrum of the underlying CR electron population and ii) the turbulence of the  magnetic field and hence of the carrying intergalactic medium. In the case of a uniform magnetic field \rev{and a power law as the CR energy distribution}, the polarisation fraction only depends on the \rev{power-law} index of this energy distribution, which can be rewritten in terms of the spectral index $\alpha$ of the emission spectrum \citep[cf. Eq. 5.46 by][]{Ginzburg:1979}:

\begin{equation}
    \pi = \frac{\alpha+1}{\alpha+5/3} \approx \rev{0.7} \quad \text{for }\alpha=\frac{1}{2}. \label{eq:pol_expectancy}
\end{equation}

\noindent \rev{It is instructive to keep this limit in mind, since it provides us with a sense of what to expect for a given CR population.} If the measured value for the fractional polarisation is significantly lower, it is likely due to turbulence in the magnetic field along the line of sight. This produces an increasingly diverse distribution of polarisation angles, which can further 'fan out' due to Faraday rotation.

In \cref{fig:polarization} we present the polarisation fraction $\pi_\nu$ of the simulated ORC at a frequency of $\nu=\SI{944}{\mega\hertz}$ given by \cref{eq:polfrac}. \rev{Faraday rotation along the line of sight is taken into account according to \cref{eq:faradayrotation}.} Each quantity ($Q_\nu, U_\nu$ and $I_\nu$) was smoothed according to the resolution of ASKAP at $\nu=\SI{944}{\mega\hertz}$ ($\theta=\SI{15}{\arcsecond}$, see \cref{subsec:radiomocks}), where we assumed the same distance as to ORC Physalis at $z=0.017$ \citep{Koribalski:2024a}. The time, scale, orientation and circular mask of the map is the same as in \cref{fig:synch_spectralindex}. For the most part, the simulated ORC has a degree of polarisation between $20-40\%$. We analysed the rotation measure for the simulation and found that it is negligible outside the halo centre \rev{(see \cref{fig:rotationmeasure}). Also, the beam size of \revb{\mbox{2.5 kpc}} adopted for the mock image is much smaller than the smoothing length \revb{of \mbox{$\sim$10 kpc}} for the gas particles at the ORC position (see \cref{subsec:pressuremap}). This means that the polarisation is dominated by emission at the shock ring. We note that magnetic field fluctuations below our resolution limit must exist and might contribute to the final polarisation. However, is is not possible to make a quantitative prediction for this particular effect based on the this simulation.}

Judging by their emission spectrum, CRs are injected quite efficiently in our simulation and yield a spectral index close to the DSA limit of $\alpha=0.5$ (see \cref{subsec:spectrum_radiopower}). The fact that we arrive at a much lower mean polarisation degree than the expected value given by \cref{eq:pol_expectancy} is an indicator for the turbulence that is present inside the shocked plasma along the line of sight, as explained above. The sporadic white spots in \cref{fig:polarization}, which show a deviantly high degree of polarisation in comparison, correspond to sight lines with singular CR populations that are not contaminated by additional emissions along the line of sight. This is mostly a numerical artefact, which is also evident from the overall 'popcorn-like' appearance of the map. Although we are already working with very high resolution here, the shock finder (which is highly resolution dependent) occasionally fails to pinpoint the shock location smoothly (see \cref{subsec:pressuremap}). Additionally, the injection of CRs strongly depends on the evaluated Mach number as well as on the estimated shock obliquity, which determines the local acceleration efficiency and can further add uncertainties to the simulated polarisation. We point out, however, that this has no significant impact on the overall value of the polarisation fraction across the shock ring. \cref{fig:polarization} therefore qualifies for a comparison with observed ORCs, which is carried out in the discussion (\cref{sec:discussion}).

\subsection{Shock indicators}
\label{subsec:pressuremap}

\begin{figure}[!t]
\centerline{
    \includegraphics[width=0.5\textwidth, trim={0 0 0 0}]{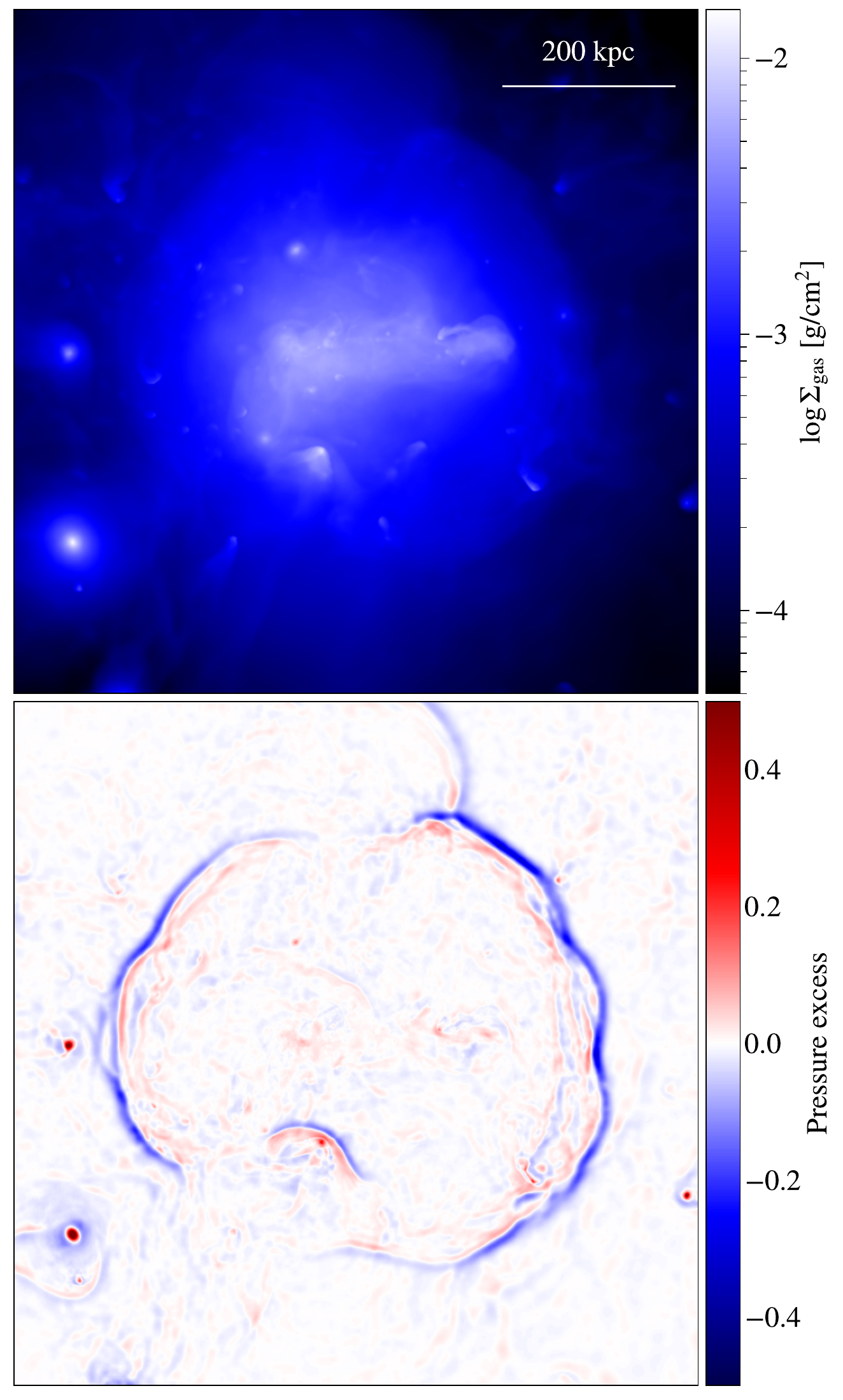}}
    \caption{Gas surface density (top panel) and shock map created with Gaussian image filtering (bottom panel, see text for details) at $t=\SI{73}{\mega\year}$. The scale in the top right corner of the density map indicates a size of \SI{200}{\kilo\parsec}.}
    \label{fig:pressureexcess}
\end{figure}

The built-in shock finder in the code \textsc{OpenGadget3} is based on a geometrical on-the-fly scheme that utilises pressure gradients to construct the shock normal along which the Rankine-Hugoniot jump conditions are evaluated. After masking cold flows and shear flows based on threshold pressure and velocity jumps, the algorithm finally inferes shock quantities such as the compression ratio and Mach number from approximated hydrodynamic quantities up- and downstream of the shock \citep{Beck:2016b}. As mentioned in \cref{subsec:polarization}, the accuracy of the shock finder strongly depends on the resolution of the simulation. Specifically, the size of the local smoothing kernel of the respective SPH particle defines the scale on which the shocks are resolved. In our case, the mean smoothing scale of the simulated ORC is of the order of $\SI{10}{\kilo\parsec}$ at $t=\SI{73}{\mega\year}$, i.e. when the size and morphology matches best to observed counterparts. For reference, this is twice as big as the smoothing kernel imitating the resolution limit of ASKAP at $\nu=\SI{944}{\mega\hertz}$, indicated by the small circle in the lower left corner of the polarisation map in \cref{fig:polarization}. This explains the appearance of the spot-like features in the inferred mock radio images -- particularly in the polarisation features -- which encompass the extent of the SPH particles in these regions.

In order to illustrate the actual extent and width of the shock region, it is instructive to compare the results with an alternative way to determine the shock position. In particular, Gaussian filtering of pressure maps has proven to be quite efficient in revealing pressure jumps that indicate a shock in the medium \citep[see also][]{Dolag:2005b,Ragagnin:2017,Sommer:2024}. Naturally, this approach is only feasible during post-processing of the simulation, since it requires a 2D projection in the correct plane and visual inspection. Nevertheless, it is a convenient tool for double-checking the conclusion of the shock finder. The process is as follows: starting from a pressure map $\mathcal{X}(x,y)$, we smoothed the image by applying a 2D Gaussian kernel function $G(x,y)\propto \exp[-(x^2+y^2)/(2\sigma^2)]$ to each pixel of the original image, yielding a blurred variant $\tilde{\mathcal{X}}(x,y)$. Sharp edges in the original map remain present, but are naturally attenuated in the new image. If we now subtract the two images from each other, features spanning over a larger scale than the smoothing kernel $\sigma$ vanish and only sharp features remain. Lastly, the contribution of each pixel was normalised by the local value of the blurred image, to ensure all edge features appear equally strong in the resulting map, no matter the pressure contrast. The final shock map is therefore given by

\begin{equation}
    \mathcal{X}_{\rm fin}(x,y) = \frac{\mathcal{X}(x,y)-\tilde{\mathcal{X}}(x,y)}{\tilde{\mathcal{X}}(x,y)}.
\end{equation}

\noindent In \cref{fig:pressureexcess} we present such a shock map of the system generated by the method explained above. The time and scale of the simulated ORC is again the same as in the polarisation map in \cref{fig:polarization}. The upper panel shows the integrated surface density of the gas, while the bottom panel displays a shock map in the same scale. The pressure at each pixel was determined via the Compton-$y$ parameter, which can be used as a proxy for the integrated electron pressure along the line of sight. Finally, we used a Gaussian smoothing kernel length of $\sigma=5$ pixels, which corresponds to approximately $\SI{6}{\kilo\parsec}$ in this case. The shock map in \cref{fig:pressureexcess} reveals clear pressure edges, with strikingly circular geometry, which was not evident in the shock finder output.

\section{Discussion}
\label{sec:discussion}

We showed in \cref{sec:results}, that merger-accelerated shocks can produce giant circular radio features around massive galaxies and galaxy groups. Their morphology closely resembles those of a recently discovered class of radio objects, so-called ORCs \citep[e.g.][]{Norris:2021,Norris:2022,Koribalski:2021,Koribalski:2024a}. It is yet under debate as to what these peculiar structures are, since they have a number of unusual traits. Particularly, they display i) a remarkably circular, edge-brightened morphology, ii) have large sizes comparable to the virial radius of a central elliptical galaxy group and iii) exhibit exceedingly high radio powers, judging by self-similar scaling of their halo mass \citep[cf. \mbox{Figure 6} by][]{Koribalski:2024a}.

\subsection{Comparison to observations and Dolag et al. (2023)}
\label{sec:comparison2asin}

The effectiveness of galaxy mergers in producing circular radio features matching ORCs was first shown in a pilot study by \cite{Dolag:2023}, where the authors conducted cosmological zoom-in simulations of a Milky-Way-like halo with a mass of approximately $10^{12}\,\Msun$. Since the simulations were performed in a pure hydrodynamics fashion without magnetic fields, they needed to estimate the radio brightness during post-processing. That was realised by adding CRs to shock regions, where their amount was chosen such that the integrated CR energy was a fixed fraction of the dissipated shock energy, based on the shock finder output at a single time step. Finally, by assuming a constant and homogeneous magnetic field, they could infer the spectral radio flux based on synchrotron emission. Although the morphology and size of ORCs could be reproduced, these simulations predicted radio powers that were far too low compared to what is measured in ORCs. However, it was unclear which assumptions specifically led to this deficiency. In this work, we address two main shortcomings of the study by simulating i) a massive galaxy group ($M\sim10^{13}\,\Msun$), comparable to observed counterparts and ii) using a magnetohydrodynamic scheme with a self-consistent, on-the-fly treatment of spectral CRs.

\begin{figure}[!t]
\centerline{\includegraphics[width=0.5\textwidth, trim={0 0 0 0}]{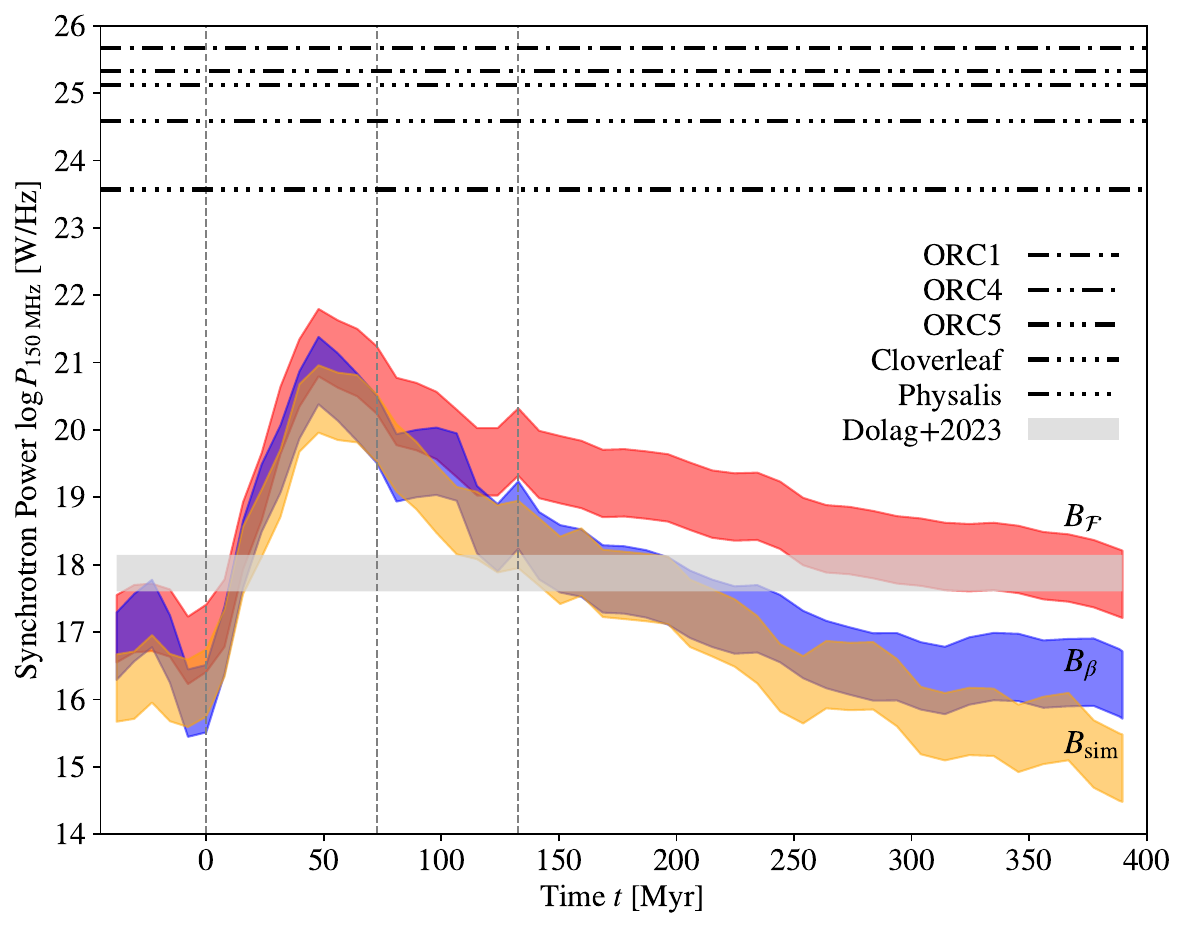}}
    \caption{Comparison of the radio luminosity profile of the simulated ORC with observations. The three coloured regions show the synchrotron power at $\nu=\SI{150}{\mega\hertz}$ inferred from three different magnetic field models (see text for details). The grey band represents the result from the simulation by \citet{Dolag:2023}, while the horizontal dashed-dotted lines indicate the level of observed counterparts at the respective frequencies: ORC1 \citep{Norris:2022}, ORC4 \citep{Norris:2021}, ORC5 \citep{Koribalski:2021}, Cloverleaf (Koribalski et al. in prep.) and Physalis \citep{Koribalski:2024a}. The vertical dashed lines mark the three time steps chosen for the visualisation of the merger sequence in \cref{fig:Xray_CRe}.}
    \label{fig:radiopower_models_ORC5}
\end{figure}

As discussed in \cref{subsec:radiomocks}, it is necessary to proceed with caution when evaluating the absolute value of the magnetic field in a cosmological simulation. It is often underestimated, because extremely high resolution is technically required to resolve the turbulent dynamo process in the intergalactic medium, which drives magnetic field amplification \citep[e.g.][]{Vazza:2014,Vazza:2018,Boess:2024,Steinwandel:2024}. Since the radio properties of the system are based on synchrotron emission, this can significantly alter the final result. To obtain more realistic predictions, it is therefore appropriate to compare the simulated magnetic field strength with analytic predictions for the system. In \cref{fig:radiopower_models_ORC5}, we show the time-dependent radio luminosity due to synchrotron emission at \SI{150}{\mega\hertz} for the simulated ORC. The three vertical lines represent the three times from the visualisation in \cref{fig:Xray_CRe}, which are each about \SI{70}{\mega\year} apart, starting shortly before the merger. The coloured regions represent three different models for the magnetic field strength, where the lower end is the prediction by the respective model and the upper end represents a boost by a factor of 10. The orange area shows the luminosity profile determined by the simulated magnetic field strength $B_{\rm sim}$, while the other two were generated by assuming a balance between i) magnetic field pressure and thermal pressure ($B_\beta$, blue) and ii) magnetic field pressure and turbulent pressure ($B_\mathcal{F}$, red). Hence, the local magnetic field strength is given by

\begin{align}
    B_{\beta} &= \sqrt{\frac{8\pi P_{\rm th}}{\beta}} \label{eq:bfield_beta}\\
    B_{\mathcal{F}} &= \mathcal{F} \sqrt{4\pi \rho v_{\rm turb}^2} \label{eq:bfield_turb}
\end{align}

\noindent respectively in the two cases, where $P_{\rm th}$ is the thermal pressure, $\rho$ the mass density and $v_{\rm turb}$ the turbulent velocity. The latter is defined as $v_{\rm turb} = \left| \mathbf{v}  - \mathbf{v}_{\mathrm{bulk}} \right|$, where $\mathbf{v}$ is the local velocity. The local bulk velocity $\mathbf{v}_{\mathrm{bulk}}$ is computed as the kernel weighted mean velocity of the neighbouring particles

\begin{equation}
    \mathbf{v}_{\mathrm{bulk},i} = \sum\limits_{j=0}^{N_\mathrm{ngb}} \frac{m_j}{\rho_j} \mathbf{v}_j W(\vert\mathbf{r_i} - \mathrm{r_j}\vert, h_i).
\end{equation}

\noindent In order to keep the models realistic overall, the value of the plasma-$\beta$ value of 20 and the balance factor $\mathcal{F}=2$ were scaled for the whole simulation such that the field strength in the central regions of the merging haloes is of the order of several $\mu\rm G$, which is in accordance with measurements for galaxies in this mass range \citep[e.g.][]{Chyzy:2004,Opher:2009,Nikiel-Wroczynski:2020}. As expected, the two models generally produce higher values for the total magnetic field strength than the simulated one, reflected by the systematically greater inferred radio power. The turbulent model $B_\mathcal{F}$ is overall the most powerful and is particularly visible at later times. \rev{At $t=\SI{73}{\mega\year}$, i.e. when we identified the structure as an ORC (see \cref{fig:synch_spectralindex}), the maximum of the field strength distribution of the respective particles is $\sim 1\,\mu\rm G$ (the histogram can be found in the Appendix in \cref{fig:bfield_strength}). Recent polarimetry measurements of \mbox{ORC J0356–4216} \citep{Taziaux:2025} report field strengths of the same order of magnitude, which are about twice as large as in our simulated ORC.}

The results from the pilot study by \citet{Dolag:2023} introduced in the beginning of this Section are visualised as the horizontal grey band in \cref{fig:radiopower_models_ORC5}, where its width represents the overall extent of the total radio power inferred in post-processing of their simulation, \rev{ranging between $4\times10^{16}\, \rm W\, Hz^{-1}$ and $1.4\times10^{17}\, \rm W\, Hz^{-1}$ at $\nu=\SI{944}{\mega\hertz}$ \citep[cf. Fig. 9 by][]{Dolag:2023}. Since their observer frequency was at a higher value than in our figure\footnote{\rev{The total luminosity for observed objects is more often published at this lower frequency. Therefore we adopt this same frequency to enable comparison with more counterparts.}}, we boost the quantities by \citet{Dolag:2023} by a factor of ten\footnote{\rev{Such scaling for synchrotron emissivity is valid for a relatively hard energy spectrum, as in our case, and can be validated by comparing the power at $10^2\,\rm MHz$ and $10^3\,\rm MHz$ in \cref{fig:radiopower_spectrum}}}.} Compared to their estimate, our simulation exceeds it by at least three orders of magnitude at the moment of peak brightness -- already when using the radio power inferred from the simulated magnetic field strength alone. But even taking the boost of stronger field models into account, we still do not reach the luminosity of observed counterparts indicated by the various dashed-dotted lines in \cref{fig:radiopower_models_ORC5}. Consequently, also the surface brightness is under-predicted by the simulation. Physalis, for example, has a flux density of the order of $10^{-1}\,\rm mJy/beam$ in the outer radio rings (\rev{shown in the Appendix in \cref{fig:physalis_map}}), while the simulated counterpart is about three orders of magnitude lower at $10^{-4}\,\rm mJy/beam$ (see \cref{fig:synch_spectralindex}). This means that, although we now resolve a halo in the correct mass regime and reproduce the expected X-ray properties (see \cref{sec:xray_props}), there is clearly still a missing aspect in the picture, that prevents us from fully explaining the physical nature of these systems.

\subsection{Potential role of AGN and stellar feedback}

Assuming no other essential process produces significant radio emission besides synchrotron emission, this lack of total radio luminosity implies insufficient relativistic CR electrons present and/or underestimated local magnetic field strengths. Since the maximum Mach number of the shock, when the simulated ORC is also the brightest, is already relatively large ($\mathcal{M}\approx5$, see \cref{subsec:spectrum_radiopower}), a stronger shock would not remedy the CR deficiency, because their acceleration efficiency via DSA would not increase substantially. A stronger magnetic field would in principle also give rise to stronger synchrotron emission. The problem is that several processes currently under-represented in our magnetic field models are presumed to contribute to the magnetic field state in the medium, for example Bell amplification in shocks or halo magnetisation by highly magnetised or laminar outflows. To test the effect of stronger magnetic fields in our simulation, we would need to increase the field strength across the whole system in order to stay self-consistent. But this would push the field strength in the central regions of the galaxies beyond actually observed values, and is therefore not an option. Additionally, we need to consider the relative importance of fossil CR populations in the halo, which would be present following prior injection events of CRs into the CGM of the merging galaxies. Typically, enough time would have passed for these old populations to cool down to mildly relativistic energies -- meaning they would no longer be radio-bright -- but not enough time for them to join the thermal pool of electrons. While a strong shock is necessary to accelerate \rev{electrons} from the Maxwellian distribution, a relatively weak shock would already suffice to push such fossil CRs back to relativistic energies again, effectively boosting the total radio brightness of the whole system \citep[e.g.][]{Caprioli:2018,Kang:2018}.

Apart from merger or accretion shocks, stellar outflows and AGN feedback are the other main sources for these fossil populations, and it can be assumed that they should play a role in the two galaxies in our simulation. But since it was performed in a non-radiative fashion, meaning that it did not include star formation, it is not possible to make definite statements about the actual contribution of stars and black holes at this point. \rev{To our knowledge, no studies exist yet about the relative importance of shocks versus stellar and AGN contributions in the energy budget of the CGM. \citet{Shabala:2024} give an expression to estimate the total energy injected into the medium by a pair of jets in their Eq. A6. Assuming that a violent AGN outburst can inflate a sphere with a radius of \mbox{200 kpc} into an ambient medium with a density of $n_0 = 10^{-2}\, \rm cm^{-3}$, this would amount to a total energy budget from this AGN episode of $E_{\rm AGN}\approx 10^{61}\, \rm erg$. Studies of clusters with halo masses \mbox{$\sim$$10^{15}\, \Msun$} indicate that major mergers can release energies of the order of $10^{64}\, \rm erg$ in these systems \citep{Feretti:2012,Dolag:2023}. We can scale this value down to group scales by assuming that this energy budget scales with the radio power versus halo mass relation, which persists across group and cluster masses. Thus, we arrive at an energy release of $E_{\rm merger}\approx10^{61}\, \rm erg$ in groups with masses around $10^{14}\, \Msun$ \citep[cf. Fig. 2 by][]{Cuciti:2023}. Hence, according to this rudimentary estimate, this would mean that the energy releases by AGN and mergers are on the same order of magnitude inside a group environment.

But even if AGN could contribute significantly more energy in relativistic electrons compared to merger events, it would not resolve the issue at hand, as becomes apparent from \cref{fig:radiopower_models_ORC5}. There, the upper end of the inferred luminosity profiles from the simulation equates to an energy boost by a factor of 10, which evidently, would not suffice to reach the observed values.} Nevertheless, the fractional polarisation actually also points towards the importance of fossil CRs in this system. The simulation produces values mainly between $20-40\%$ (see \cref{subsec:polarization}), while measurements of \mbox{ORC~1} \rev{and \mbox{ORC J0356–4216}} report lower values mostly between $10-20\%$ in the interior and reaching $\sim$30\% at most in the ring \citep{Norris:2022,Taziaux:2025}. If there are significant amounts of old CR populations present, one would certainly increase the number of emission centres along the line of sight compared to the case where electrons were accelerated from the thermal pool only. But since also the direction of the magnetic field vector varies along the line of sight—and the polarisation angle depends on that direction—this would naturally diversify the distribution of polarisation angles present in the total emission. Therefore, the total polarisation fraction along a given line of sight would decrease with additional fossil CRs populations, which is necessary to agree with observations. Thus, in order to provide a concluding answer to the question whether fossil CR populations are indeed the missing key to explaining the radio features of ORCs, one would need to perform the simulation again with star formation and proper feedback implementation, including self-consistent CR treatment. This remains a challenge for current state-of-the-art cosmological codes and will be addressed in future work.

\rev{
\subsection{Outlook on other formation scenarios}

In addition to the merger shock paradigm, a variety of other formation pathways has been introduced in the literature so far. Among these, only two models have been tested via simulations as well—(i) the AGN driven model and (ii) the starburst wind model.

The most recent study on the AGN model was conducted by \citet{Shabala:2024}, where the authors presented hydrodynamical simulations of an AGN jet, where the synchrotron emissivity was estimated in post-process. \citet{Shabala:2024} argued that AGN induced material would rise and evolve too slowly in comparison with the energy loss timescale of embedded CRs, meaning that AGN ejections alone cannot explain the extent and luminosity of ORCs. In order to examine the feasibility of CR re-acceleration, they set up a shock to traverse through the existing AGN bubble and obtained promising results that produced correct morphologies. Also as an idealized hydrodynamical simulation, \citet{Coil:2024} recently presented a study on the starburst wind model, where the authors analysed the shocks induced by a sudden wind within an isolated galaxy and reported shock morphologies extending over several hundred of kiloparsecs.

Since both these models produce correct sizes, the next step would be to quantify the total radio power that can be reached via these formation pathways and compare it to observed counterparts. In particular, it is imperative to obtain a polarisation estimate from magnetohydrodynamic simulations of both models, as this property can change significantly depending on the underlying physical processes—and as such might be essential in estimating the likelihood of different formation scenarios.
}

\section{Summary and conclusion}
\label{sec:summary}

In this article, we presented a first of its kind zoom-in simulation of a galaxy group performed in a magnetohydrodynamic setup with spectral CR treatment on-the-fly. The scientific aim was to test the limits of a strong contender for a formation channel for ORCs -- a new class of radio objects whose origin is still a puzzle. The idea is that during a major merger event between two galaxies, the immense energy dissipation can release several strong bow shocks (i.e. merger shocks). If the projection is close to perpendicular to the merger axis, these shocks can form a giant ring, that closely resembles the edge-brightened radio features now identified as ORCs in observations.

This simulation incorporates two major improvements over the pilot study by \citet{Dolag:2023}, who introduced this formation path: i) resolving a halo ten times more massive, with a final virial mass of about $10^{13}\,\Msun$, now in agreement with the group mass measured at the centre of ORCs, and ii) determining the CR population and local magnetic field direction self-consistently, rather than in post-processing. We showed that the simulation reproduces the X-ray surface brightness of a galaxy group in that mass range and demonstrated that the peculiar offset between the X-ray and radio-bright centres observed in ORC Physalis \citep{Koribalski:2024a} can be explained by the difference in gas density at the centres of the merging galaxies (\cref{sec:xray_props}). By providing mock observations of the radio surface brightness and spectral index based on working frequencies of LOFAR, GMRT and ASKAP, we demonstrated that the morphology and size agree well with observations (\cref{subsec:radiomocks}). The emission spectrum of the radio shells indicates a sonic Mach number of $\mathcal{M}\approx4.8$, which is comparable to strong shock events in massive galaxy clusters (\cref{subsec:spectrum_radiopower}). The simulation produces a fractional polarisation degree of $20-40\%$ across the radio ring (\cref{subsec:polarization}). The pressure excess map provided in \cref{subsec:pressuremap} reveals the sharp shock boundaries responsible for the radio features. 

Compared to observations, the simulation under-predicts the total radio luminosity and over-estimates its polarisation degree (\cref{sec:discussion}). Both issues would be alleviated by factoring in additional CR sources beyond DSA, the only acceleration mechanism in the present simulation. Such mechanisms would be injection of cosmic rays via stellar or AGN feedback, and could \rev{substantially} contribute to the total energy budget and boost the radio brightness when re-accelerated by the merger shock. \rev{In addition, the simulation might underestimate the local magnetic field strength at the shock, leading to a significant under-prediction of inferred synchrotron intensity.} We will implement descriptions of fossil CRs \rev{and magnetic field amplification mechanisms} in future work to investigate whether they could explain the observed radio power of ORCs.

\begin{acknowledgements}
 \rev{We thank the anonymous referee for their helpful comments and suggestions, which greatly contributed to the quality of this work.} AI, KD and IK acknowledge support by the COMPLEX project from the European Research Council (ERC) under
 the European Union’s Horizon 2020 research and innovation program grant agreement ERC-2019-AdG 882679.
 LMB is supported by NASA through grant 80NSSC24K0173.
 The calculations for the hydrodynamical simulations were carried out at the Leibniz Supercomputer Center (LRZ) under the project \rev{pn69va and with support of the Computational Center for Particle and Astrophysics (C2PAP)}. The following software was used for this work: \textsc{julia} \citep{bezanson+17:julia}, \textsc{Matplotlib} \citep{hunter:2017:matplotlib}, \textsc{Smac} \citep{Dolag:2005}.
\end{acknowledgements}

\bibliographystyle{style/aa_url}
\bibliography{bib}

\begin{appendix}

\rev{
\section{\rev{Mach number distribution}}

\begin{figure}[!h]
\centerline{
    \includegraphics[width=0.5\textwidth, trim={0 1cm 0 0}]{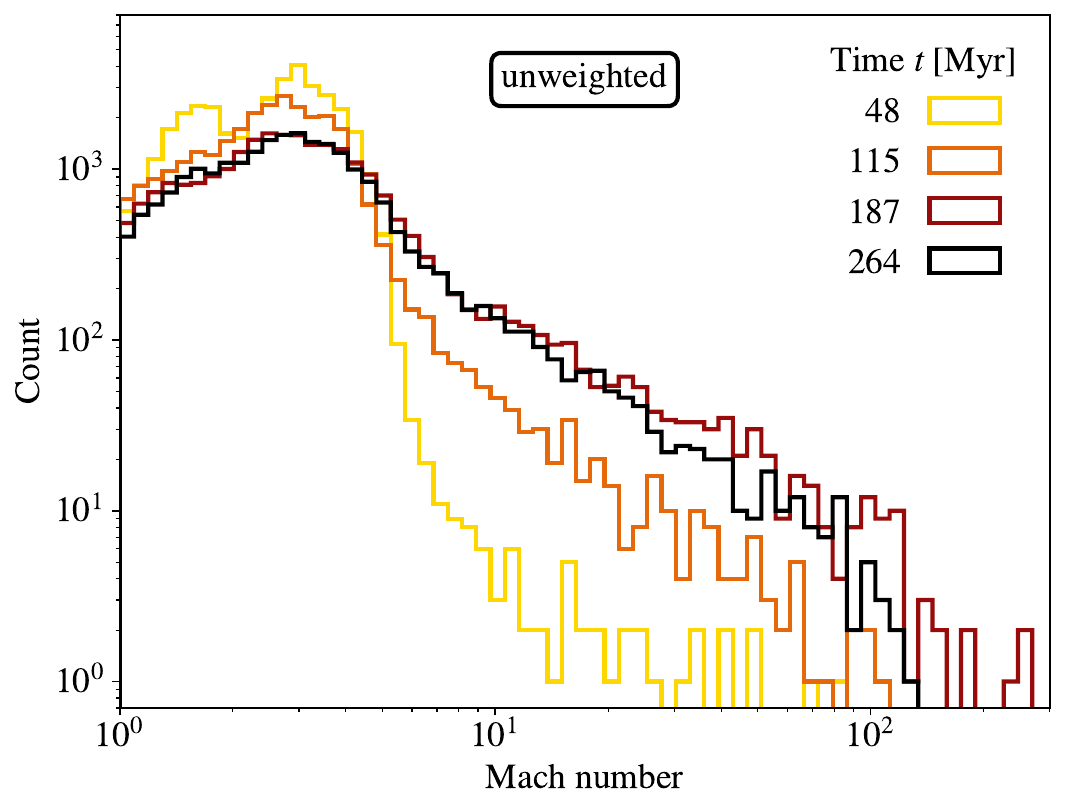}}
    \centerline{
    \includegraphics[width=0.5\textwidth, trim={0 1cm 0 0}]{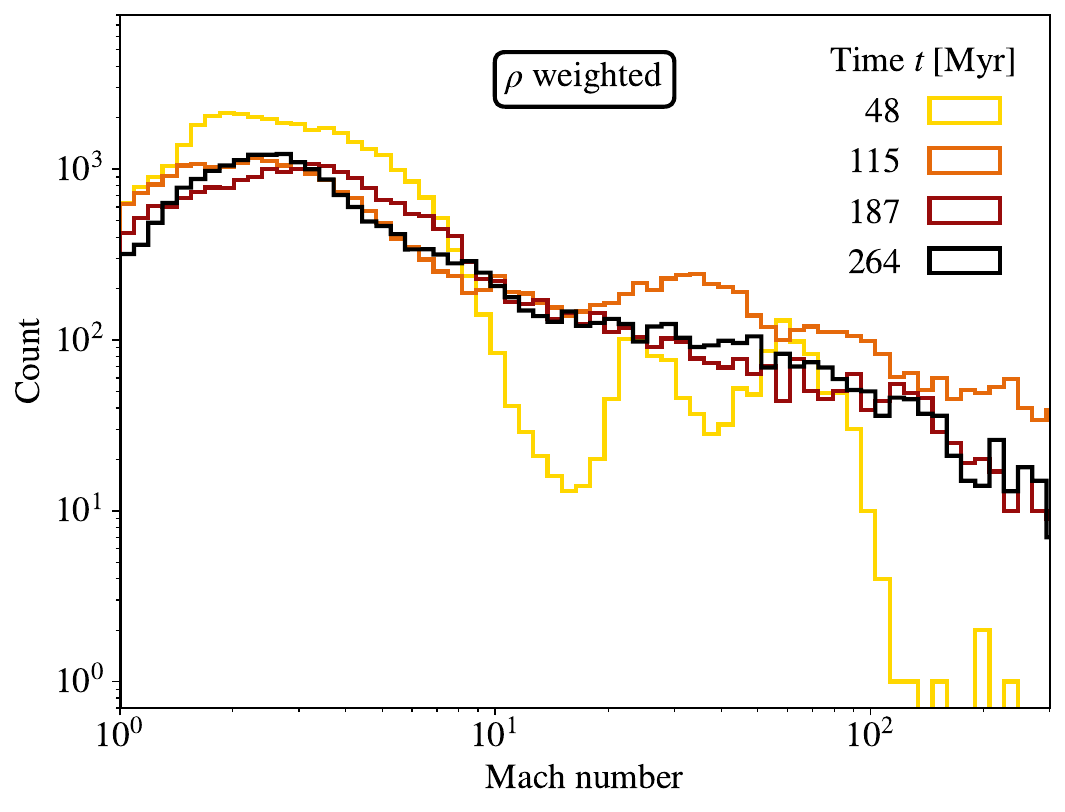}}
    \centerline{
    \includegraphics[width=0.5\textwidth, trim={0 0 0 0}]{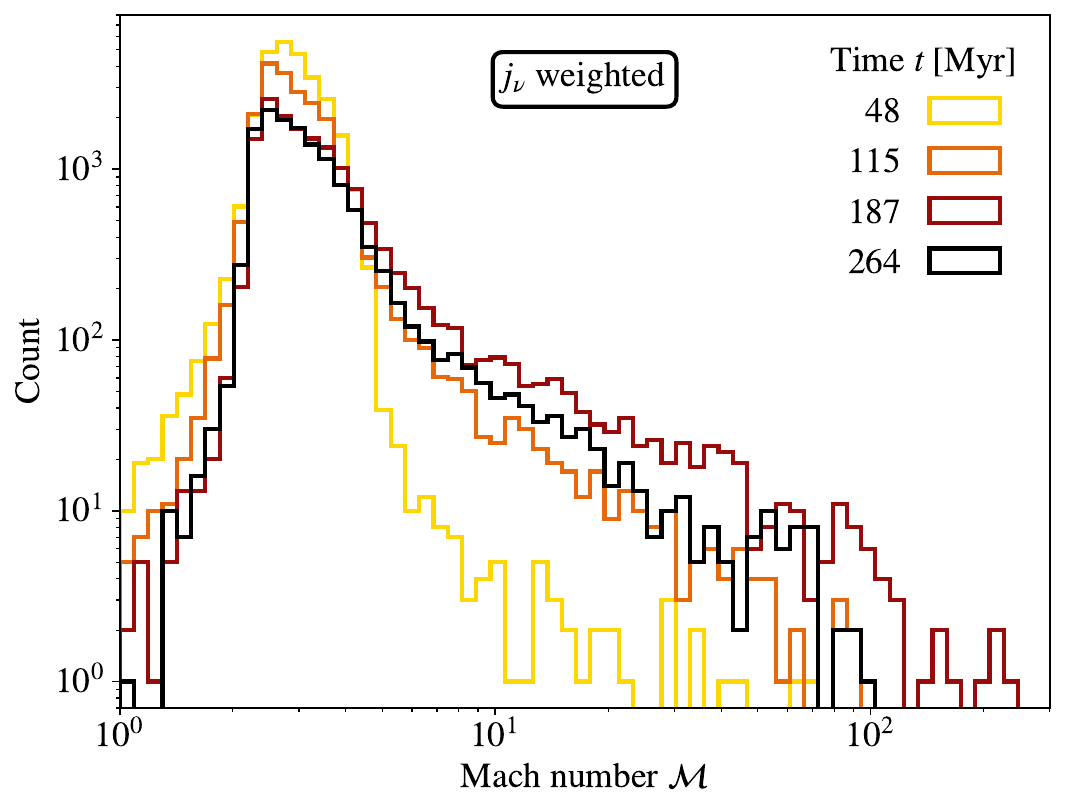}}
    \caption{\rev{Mach number distribution of the simulated ORC. Top: Unweighted distribution. Middle and bottom: distribution weighted by the local density and synchrotron emissivity, respectively. The colour gradient indicates the four successive times matching the spectrum evaluation moments to the right of \cref{fig:radiopower_spectrum}.}}
    \label{fig:machdistribution}
\end{figure}

We show in \cref{fig:machdistribution} the Mach number $\mathcal{M}$ distribution for the gas particles which make up the shock structure identified as an ORC in the simulation. For the same particles, we showed the time-dependent radio power and synchrotron emission spectrum in \cref{fig:radiopower_spectrum}. The three panels represent the distribution with three different weightings: unweighted (top), density weighted (middle) and synchrotron emissivity weighted (bottom). In each case, the distribution is dominated by contributions below $\mathcal{M}=10$, with a tail towards high Mach numbers. The broadness of the bulk distribution varies with the weighting method, while the maximum of the distribution stays relatively constant around $\mathcal{M}\approx3$.
}

\section{Rotation measure}
\label{app:rotmeasure}

\begin{figure}[!h]
\centerline{
    \includegraphics[width=0.5\textwidth, trim={0 0 0 0}]{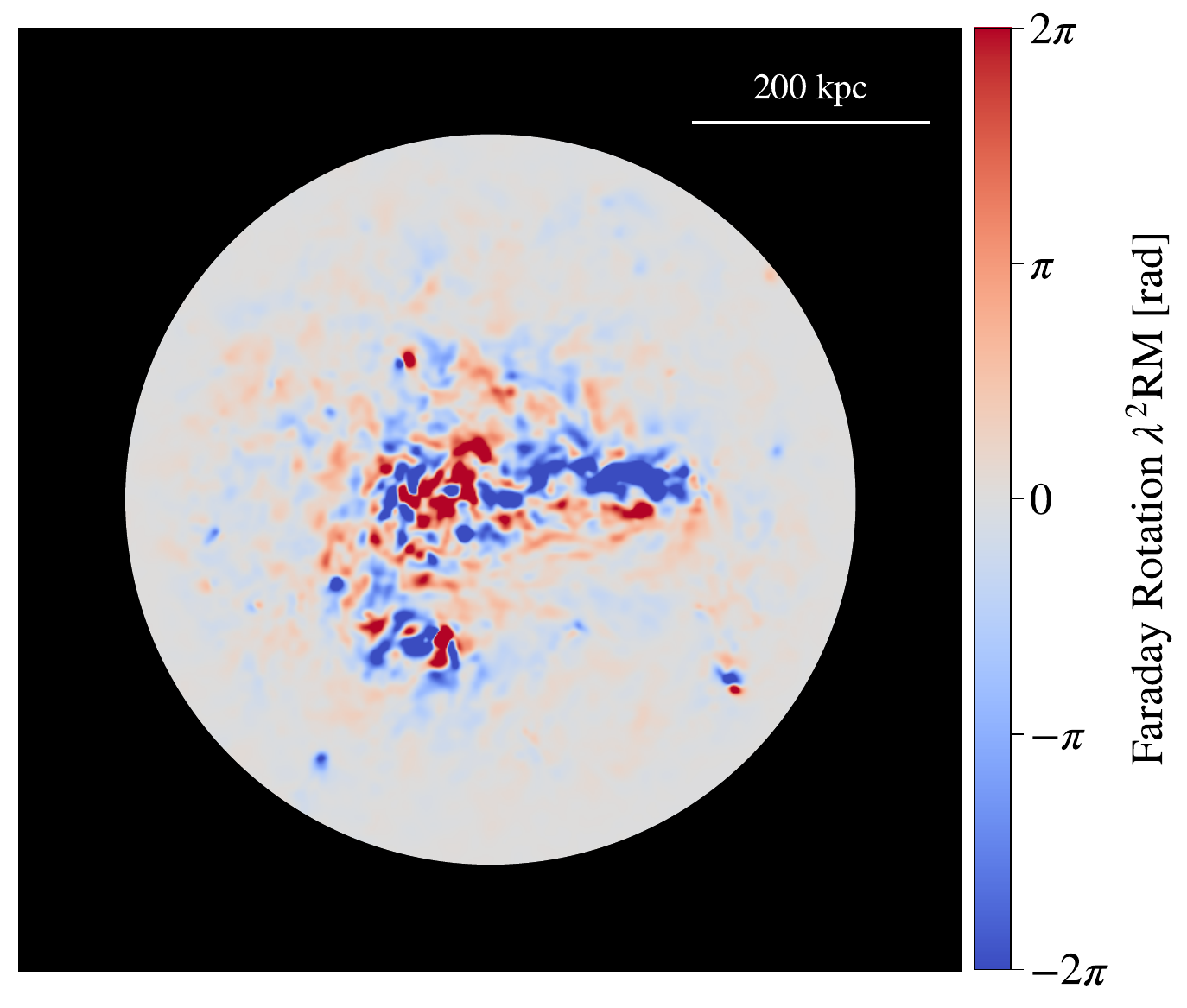}}
    \caption{Faraday rotation of the polarisation angle (at 944 MHz) along the line of sight at $t=\SI{73}{\mega\year}$ in the projection perpendicular to the merger axis.}
    \label{fig:rotationmeasure}
\end{figure}

\cref{fig:rotationmeasure} shows the correction to the polarisation angle $\chi$ due to Faraday rotation along the line of sight (see \cref{eq:faradayrotation}) of the simulated halo. We assume a frequency of $\nu=944\, \rm MHz$ which corresponds to a wavelength of $\lambda=0.32\,\rm m$. \rev{While the central region displays significant Faraday rotation, the outskirts at radial distances of about $200\,\rm kpc$ -- where also the ORC is situated -- have rotation angles close to 0. Depolarisation of the signal from the radio ring is hence only caused by turbulence of the emission centres along the line of sight and not by successive Faraday rotation.} This is mostly due to the much smaller gas density at these radii. Therefore, Faraday rotation of synchrotron emission plays only a minor role in determining the polarisation fraction at the shock front (see \cref{subsec:polarization}).

\section{Entropy features}
\label{app:entropyexcess}

\begin{figure}[!h]
\centerline{
    \includegraphics[width=0.5\textwidth, trim={0 0 0 0}]{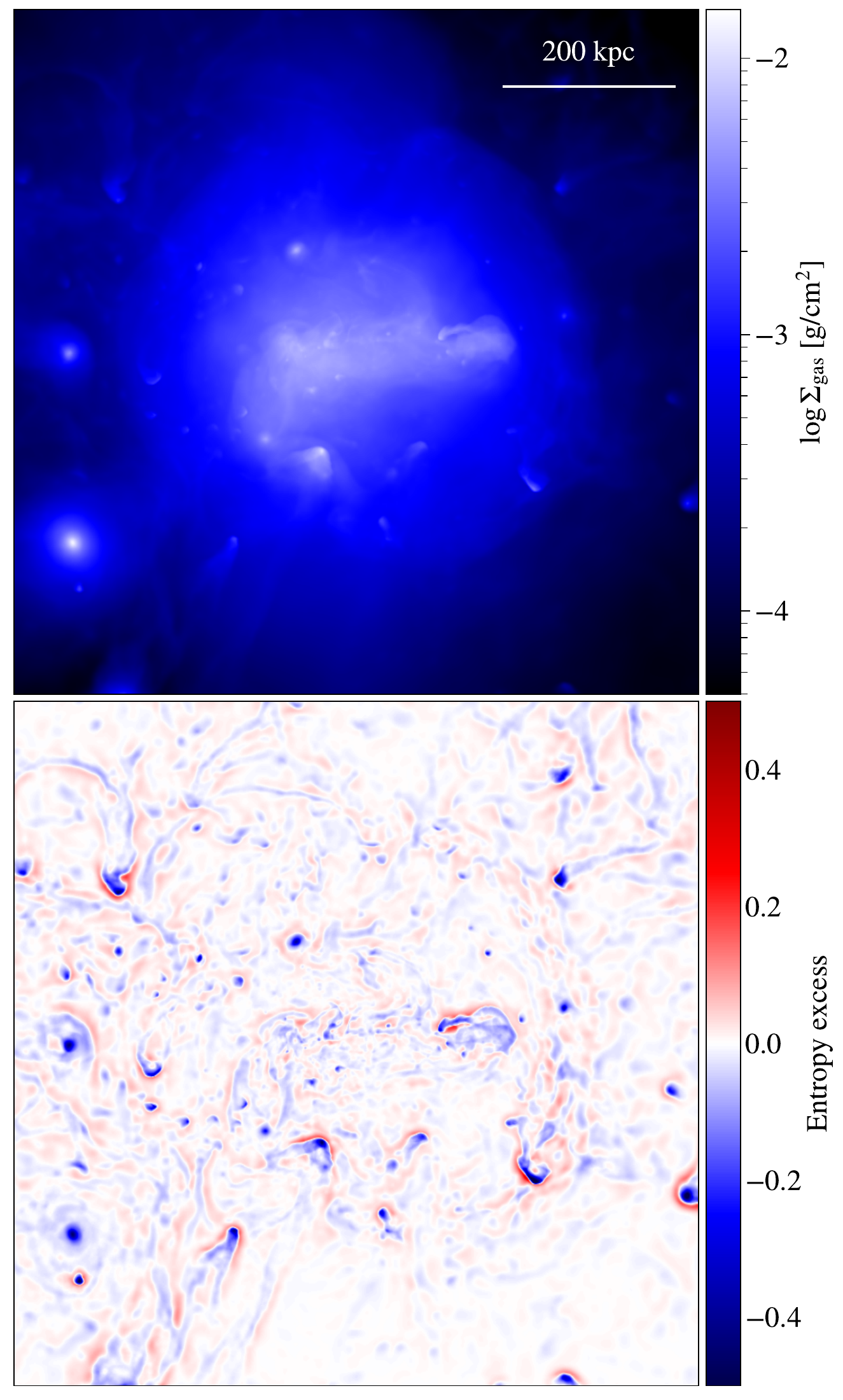}}
    \caption{Gas surface density (top panel) and entropy map created with Gaussian image filtering (bottom panel, see \cref{subsec:pressuremap} for details) at \mbox{$t=\SI{73}{\mega\year}$}. The scale in the top-right corner of the density map indicates a size of \SI{200}{\kilo\parsec}.}
    \label{fig:entropyexcess}
\end{figure}

Applying the same technique as described in \cref{subsec:pressuremap}, we show a projected entropy map of the simulated system in \cref{fig:entropyexcess}, where the projection and time is again the same as of the pressure map in \cref{fig:pressureexcess}. As a proxy for the entropy $S$ we use

\begin{equation}
    S = \frac{T}{\rho^{2/3}},
\end{equation}

\noindent where $T$ and $\rho$ are the local temperature and density of the gas, respectively. No coherent global structures are visible. Instead, the system displays a random entropy excess pattern, which is indicative of the turbulent motion in action inside the circumgalactic medium. It is mostly driven by the small satellite structures orbiting within the group halo.

\rev{
\section{Magnetic field strength}
\label{app:bfieldstrength}

\begin{figure}[!h]
\centerline{
    \includegraphics[width=0.5\textwidth, trim={0 0 0 0}]{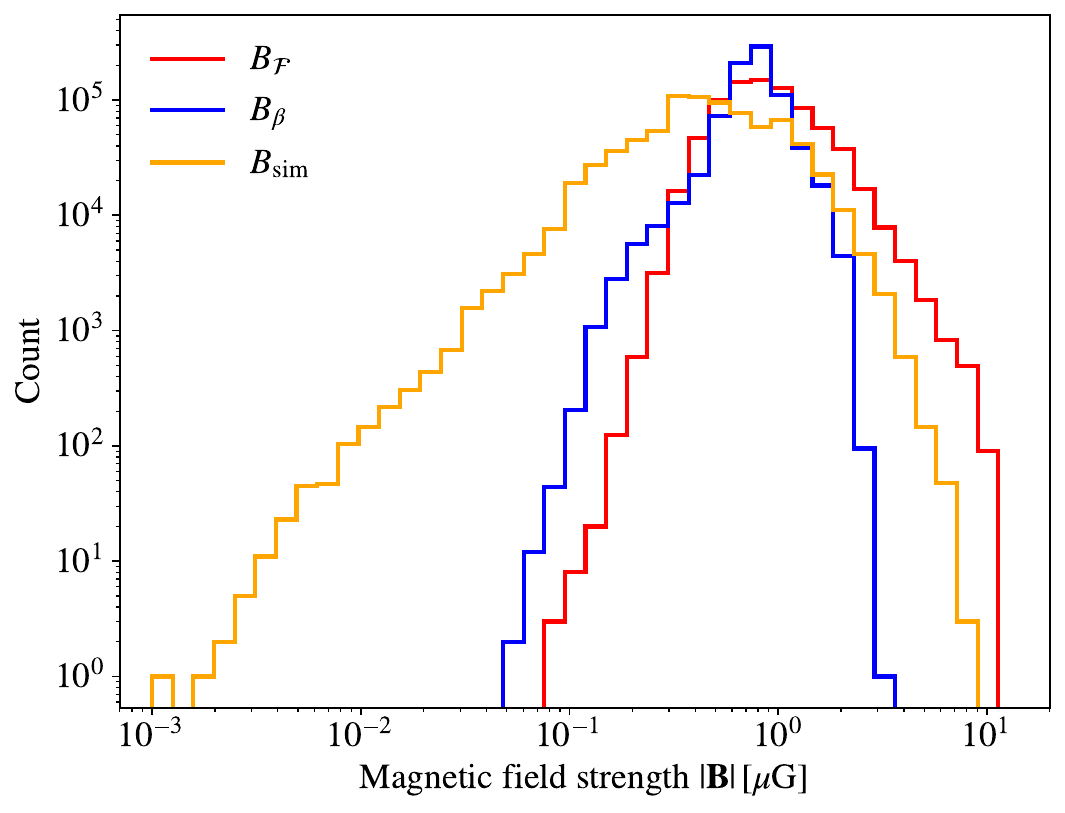}}
    \caption{\rev{Histogram of magnetic field strength of all particles inside simulated ORC at $t=\SI{73}{\mega\year}$. The blue and red lines show the distribution employing equipartition arguments (see \cref{eq:bfield_beta,eq:bfield_turb}), while the yellow line represents the simulated magnetic field strength.}}
    \label{fig:bfield_strength}
\end{figure}

In \cref{fig:bfield_strength} we show the distribution of magnetic field strengths inside the simulated ORC. The three magnetic field models employed in this figure were introduced in \cref{sec:comparison2asin}. It is clearly visible how the tail of low field strengths disappears when moving from the simulated magnetic field $B_{\rm sim}$ to either equipartition model, $B_{\mathcal{F}}$ or $B_{\beta}$. The equipartition models also shift the maximum of the distribution towards a higher value, which is then of the order of a $\mu$G. The turbulent field model $B_{\mathcal{F}}$ -- which was employed in the calculation of the synchrotron emissivity throughout the analysis of this paper -- produces overall the highest field strengths.
}

\rev{
\section{Physalis flux map}

\begin{figure}[!t]
\centerline{\includegraphics[width=0.45\textwidth, trim={0 0 0 0}]{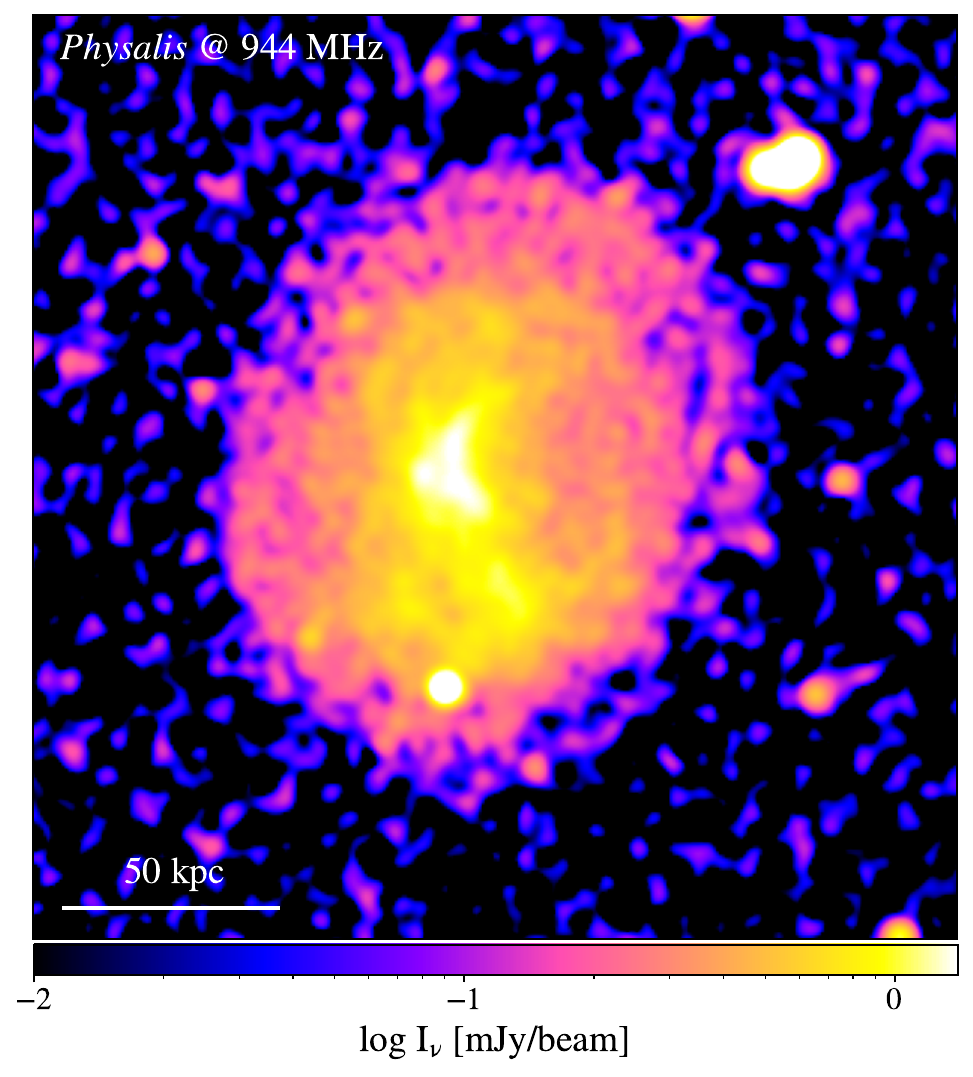}}
    \caption{\rev{Radio surface brightness of ORC Physalis at \SI{944}{\mega\hertz} \citep{Koribalski:2024a}. The scale in the bottom-left corner represents \SI{50}{\kilo\parsec} at the measured distance of $z=0.017$.}}
    \label{fig:physalis_map}
\end{figure}

In \cref{fig:physalis_map} we show the radio flux image of the observed ORC Physalis, used for comparison with our simulation throughout our work. The data cube was kindly provided by \citet{Koribalski:2024a}.

}
\end{appendix}

\end{document}